\documentclass[traditabstract]{aa}
\usepackage{verbatim} 
\usepackage{amsmath}
\usepackage{amssymb}
\usepackage{graphicx}
\usepackage{booktabs}
\usepackage{xspace}
\usepackage{siunitx}
\usepackage{mathrsfs}
\usepackage{placeins}

\makeatletter

\usepackage{txfonts}
\usepackage{enumitem}
\usepackage{natbib,twoopt}
\usepackage[breaklinks=true,colorlinks=true,linkcolor=blue,urlcolor=blue,citecolor=blue]{hyperref} 
\bibpunct{(}{)}{;}{a}{}{,}             
\makeatletter
  \newcommandtwoopt{\citeads}[3][][]{\href{http://adsabs.harvard.edu/abs/#3}%
    {\def\hyper@linkstart##1##2{}%
     \let\hyper@linkend\@empty\citealp[#1][#2]{#3}}}
  \newcommandtwoopt{\citepads}[3][][]{\href{http://adsabs.harvard.edu/abs/#3}%
    {\def\hyper@linkstart##1##2{}%
     \let\hyper@linkend\@empty\citep[#1][#2]{#3}}}
  \newcommandtwoopt{\citetads}[3][][]{\href{http://adsabs.harvard.edu/abs/#3}%
    {\def\hyper@linkstart##1##2{}%
     \let\hyper@linkend\@empty\citet[#1][#2]{#3}}}
  \newcommandtwoopt{\citeyearads}[3][][]%
    {\href{http://adsabs.harvard.edu/abs/#3}
    {\def\hyper@linkstart##1##2{}%
     \let\hyper@linkend\@empty\citeyear[#1][#2]{#3}}}
\makeatother

\newcommand{\msun}{{\rm M}_{\odot}}

\newcommand{\rsun}{{\rm R}_{\odot}}

\titlerunning{Merger seismology}
\authorrunning{J.~Henneco et al.}
\makeatother

\begin{document}
\title{Merger seismology: distinguishing massive merger products from genuine single stars using asteroseismology}
\author{
J.~Henneco\inst{\ref{HITS},\,\ref{UNIHD}}\thanks{jan.henneco@protonmail.com}, F.\,R.\,N.\,~Schneider\inst{\ref{HITS},\,\ref{ZAHARI}}, S.~Hekker\inst{\ref{HITS},\,\ref{ZAHLSW}},  \and C.~Aerts\inst{\ref{IVS},\,\ref{RUN},\,\ref{MPIA}}
}

\institute{Heidelberg Institute for Theoretical Studies, Schloss-Wolfsbrunnenweg 35, 69118 Heidelberg, Germany\label{HITS}
\and Universit\"at Heidelberg, Department of Physics and Astronomy, Im Neuenheimer Feld 226, 69120 Heidelberg, Germany\label{UNIHD}
\and Zentrum f\"ur Astronomie, Astronomisches Rechen-Institut (ZAH/ARI), Heidelberg University,  M\"{o}nchhofstr. 12-14, 69120 Heidelberg, Germany\label{ZAHARI}
\and Zentrum f\"ur Astronomie, Landessternwarte (ZAH/LSW), Heidelberg University, K\"{o}nigstuhl 12, 69117 Heidelberg, Germany\label{ZAHLSW}
\and Institute of Astronomy, KU Leuven, Celestijnenlaan 200D, 3001 Leuven, Belgium\label{IVS}
\and Department of Astrophysics, IMAPP, Radboud University Nijmegen, PO Box 9010, 6500 GL, Nijmegen, The Netherlands\label{RUN}
\and Max Planck Institute for Astronomy, K\"{o}nigstuhl 17, 69117, Heidelberg, Germany\label{MPIA}}
\date{Received xxxx / Accepted yyyy}
\abstract{Products of stellar mergers are predicted to be common in stellar populations and can potentially explain stars with peculiar properties. When the merger occurs after the initially more massive star has evolved into the Hertzsprung gap, the merger product may remain in the blue part of the Hertzsprung-Russell diagram for millions of years. Such objects could, therefore, explain the overabundance of observed blue stars, such as blue supergiants. However, it is currently not straightforward to distinguish merger products from genuine single stars or other stars with similar surface diagnostics. In this work, we make detailed asteroseismic comparisons between models of massive post-main-sequence merger products and genuine single stars to identify which asteroseismic diagnostics can be used to distinguish them. In doing so, we develop tools for the relatively young field of merger seismology. Genuine single stars in the Hertzsprung gap are fully radiative, while merger products have a convective He-burning core and convective H-burning shell while occupying similar locations in the Hertzsprung-Russell diagram. These major structural differences are reflected in lower asymptotic period spacing values for merger products and the appearance of deep dips in their period spacing patterns. Our genuine single-star models with masses above roughly 11.4 solar masses develop short-lived intermediate convective zones during their Hertzsprung gap evolution. This also leads to deep dips in their period spacing patterns. Because of the lack of a convective core, merger products and genuine single stars can be distinguished based on their asymptotic period spacing value in this mass range. We perform the comparisons with and without the effects of slow rotation included in the pulsation equations and conclude that the two types of stars are seismically distinguishable in both cases. The observability of the distinguishing asteroseismic features of merger products can now be assessed and exploited in practice.}
\keywords{Asteroseismology -- Methods: numerical -- Stars: oscillations -- Stars: binaries -- Stars: massive -- Stars: evolution}
\maketitle
%
%
%
%
\section{Introduction}\label{sec:introduction}

Stellar mergers occur frequently in our Universe \citep{podsiadlowski_presupernova_1992,sana_binary_2012,de_mink_incidence_2014} and are driven by a plethora of mechanisms \citep{henneco_contact_2024}. Merger products, the stars left behind after the stellar merger events, are expected to have peculiar properties. These properties include large-scale surface magnetic fields \citep{Ferrario2009,Wickramasinghe2014,schneider_stellar_2019}, peculiar chemical compositions (e.g. $\alpha$-rich young stars, \citealt{chiappini_young_2015,martig_young_2015,izzard_binary_2018,hekker_origin_2019}), peculiar rotation rates \citep{schneider_stellar_2019,wang_stellar_2022}, B[e] emission features \citep{podsiadlowski_massive_2006,wu_art_2020}, and masses above a cluster's main sequence (MS) turn-off mass (blue stragglers, e.g. \citealt{rasio_minimum_1995,sills_evolution_1997,sills_evolution_2001,mapelli_radial_2006,glebbeek_evolution_2008-1,ferraro_dynamical_2012,schneider_evolution_2015}).\\

Single-star evolution predicts that post-main-sequence (post-MS) intermediate- and high-mass stars (stars with convective cores during core-hydrogen burning, i.e. initial masses ${\gtrsim}1.3\,\msun$) evolve relatively quickly from the blue to the red side of the Hertzsprung-Russell diagram (HRD), crossing the Hertzsprung gap (HG) and becoming red supergiants (RSGs). Even though the HG crossing time depends on the assumptions for \(\text{(semi-)}\)convective mixing \citep{kaiser_relative_2020,sibony_impact_2023}, HG stars are expected to be rare. However, blue post-MS stars, most notably blue supergiants (BSGs), are abundant \citep{Castro2014,Castro2018,deBurgos2023,Bernini-Peron2023}. This is known as the blue supergiant problem. Several potential solutions to this problem exist. Extra mixing on the main sequence (MS) can enhance core masses, radii, and luminosities, while the stars reach cooler effective temperatures before the terminal-age main sequence (TAMS) \citep{Brott2011,johnston2019,kaiser_relative_2020,Johnston2021}. Another way to populate the blue part of the HG is through the blue loop phase, in which RSGs evolve towards hotter temperatures and appear as BSGs \citep{Saio2013,ostrowski_pulsations_2015}. However, these single-star evolutionary channels do not produce the full population of observed BSGs \citep{Bellinger2024}. The products of post-MS stellar mergers, that is, mergers that occur after one of the components has left the MS, also appear as BSGs \citep{Hellings1983,Hellings1984,podsiadlowski_merger_1990,podsiadlowski_progenitor_1992,morris_triple-ring_2007,Claeys2011,Vanbeveren2013,Justham2014,Menon2017,menon_evidence_2024}. Furthermore, the variety in pre-merger systems, such as the mass ratio and the evolutionary stage at which the merger occurs, naturally explains the variety of luminosities and effective temperatures of the observed BSGs.\\

Using 1D merger models, \citet{menon_evidence_2024} show that the surface diagnostics such as luminosity, effective temperature, surface gravity, and N/O and N/C ratios of post-MS merger products differ significantly from those of genuine single HG stars. Moreover, these surface diagnostics agree well with those determined from a sample of nearly 60 BSGs observed in the LMC. In their proof-of-principle study, \citet{Bellinger2024} show the potential of asteroseismology to distinguish MS stars with oversized cores because of extra mixing and core-helium burning (CHeB) stars with undersized cores among the population of BSGs. Such CHeB stars with undersized cores can be the result of post-MS mergers or a blue loop phase. These stars have significantly different internal structures, which is reflected in the mean spacing between their oscillation modes. To distinguish between post-MS merger products and genuine single HG stars, \citet{Bellinger2024} propose to search for a temporal change in oscillation frequencies, which could be detectable for the faster evolving genuine single HG stars. For lower-mass stars, \citet{rui_asteroseismic_2021} found that one can distinguish post-MS merger products from genuine single red giant branch (RGB) stars based on the mean spacing between the oscillation modes and mass estimates from other asteroseismic and surface diagnostics. One condition is that the RGB star needs to have a degenerate core before the merger. Other studies have explored the influence of mass accretion and mass loss on the seismic signals of red giant \citep{Deheuvels2022,Li2022} and more massive B-type \citep{Wagg2024} stars in binary systems.\\

Here, we focus on genuine and candidate merger product BSGs for the mass range between about 5\,M$_\odot$ and 20\,M$_\odot$. In this current era of high-cadence space photometry, multiperiodic nonradial oscillations have been detected in several BSGs in the envisioned mass regime, although the number of stars with firm detections remains  
limited compared to other classes of pulsators \citep[cf.\,][for a recent review]{Kurtz2022}. Multiple isolated oscillation modes have been detected in a set of about 40 BSGs with the ESA Hipparcos satellite \citep{Lefever2007} and in specific BSGs with the Microvariability and Oscillations of Stars \citep[MOST,][]{Walker2003} mission by \citet{Saio2006,Rigel2012a,Rigel2012b}, the Convection, Rotation and planetary Transits \citep[CoRoT,][]{Auvergne2009} space telescope by \citet{Aerts2010b}, the \textit{Kepler}/K2 \citep{Koch2010} missions by \citet{Aerts2017,Aerts2018}, and the Transiting Exoplanet Survey Satellite \citep[TESS,][]{Ricker2016} by \citet{2023Sanchez}. 

All of the above detections concern individual isolated frequencies in the amplitude spectra. However, BSG variability is more diverse than this. \citet{Bowman2019} and \citet{Ma2024} established global low-frequency power excess in a sample of about 180 K2 or TESS BSGs, several of which in the LMC. For many of these targets, \citet{Bowman2019} stressed that the low-frequency power excess occurs in addition to many isolated significant frequencies with higher amplitude. In his review of massive star variability, \citet{Bowman2023} pointed out that the origin of the low-amplitude low-frequency variability points to a spectrum of internal gravity waves triggered by core convection but is not yet firmly established \citep[see also][for interpretations in terms of sub-surface convection for the aspect of the power excess]{Cantiello2021}. \citet{Ma2024} used a BSG merger model from  \citet{Bellinger2024} to explore two not mutually exclusive physical origins of the observed frequency spectra, namely sub-surface convective motions and internal gravity waves excited by the thin convective layer connected to the iron opacity bump in the envelope. They find waves to be the more plausible explanation for the overall observed variability frequency spectra but admit that more 3D simulations are needed to come to firm conclusions.\\

In this work, we make an in-depth model-by-model comparison between the asteroseismic predictions for post-MS merger products and genuine single HG stars based on 1D stellar structure models. We focus on stars with masses between $7.8$ and $15.3\,\msun$ at similar locations in the HRD and explore which asteroseismic diagnostics help us distinguish merger products from genuine single stars. With this mass range, we include stars below and inside the mass range considered in previous works (e.g. \citealt{Bellinger2024}). In this mass range, we also avoid the added complexity that stellar wind mass loss might have on the photometric signal \citep{Krticka2018}.\\

This paper has the following structure. In Sect.\,\ref{sec:diagnostics}, we summarise the key concepts and diagnostics required for the asteroseismic characterisation of pulsating stars. We describe our computational setup for the equilibrium stellar structure models and stellar oscillation calculations in Sect.\,\ref{sec:methods}. Section \ref{sec:results} covers the results of our comparison between merger products and genuine single stars. Lastly, in Sect.\,\ref{sec:discussion} we discuss these results and draw our conclusions.

\section{Asteroseismic diagnostics}\label{sec:diagnostics}

Asteroseismology has shown that intermediate-mass stars born with a convective core have quasi-rigid rotation throughout their MS and 
undergo efficient yet poorly understood angular moment loss once beyond the MS \citep{Aerts2019,aerts_probing_2021}. From the results of the 3D merger simulations from \citet{schneider_stellar_2019} and the 1D follow-up study by \citet{schneider_long-term_2020}, we expect merger products to be slow rotators. Moreover, as shown by \citet{wang_stellar_2022}, merger products' slow rotation can explain the blue MS band in young stellar clusters. We thus expect slow rotation for both genuine HG stars and post-MS merger products.
In first instance, it is therefore justified to ignore rotation in the equilibrium models used to solve the pulsation equations, as is common practice for post-MS stars. Oscillation modes in slowly- and non-rotating stars are described with spherical harmonics $Y_{\ell}^{m}$ \citep{aerts_asteroseismology_2010}. The functional form of these spherical harmonics depends on the spherical degree $\ell$ (number of nodal lines on the surface) and the azimuthal order $m$ (number of nodal lines that cross the equator, $|m| \leq \ell$). The third quantum number required to describe oscillation mode morphologies is the radial order or overtone $n$, corresponding to the number of nodal surfaces or nodes in the radial direction in the stellar interior. This work focuses on nonradial oscillations, which are those with $\ell > 0$.  

Further characterisation of stellar oscillations depends on their restoring force. In non-rotating stars, these are the pressure gradient and buoyancy force \citep{aerts_probing_2021}. Acoustic modes, with the pressure gradient as their dominant restoring force, are called pressure (p) modes. Those with buoyancy as their main restoring force are called gravity (g) modes. Both types of modes are restricted to their respective mode cavities, determined by the (linear) Lamb frequency $\Tilde{S}_{\ell}$ and the (linear) Brunt-V\"ais\"al\"a (BV) or (linear) buoyancy frequency $\Tilde{N}$, which are defined as\footnote{By default, we use linear frequencies in this work. Therefore, we use the linear definitions of the BV and Lamb frequencies, indicated by a tilde. They are related to their angular forms $N$ and $S_{\ell}$ by $N = 2\pi\Tilde{N}$ and $S_{\ell} = 2\pi\Tilde{S}_{\ell}$, respectively. Angular frequencies are explicitly referred to as such.} \citep{aerts_asteroseismology_2010}
\begin{equation}
    \Tilde{S}_{\ell}^2 = \frac{\ell(\ell+1) c_{\mathrm{s}}^2}{4\pi^{2}r^2}
\end{equation}
and
\begin{equation}
    \Tilde{N}^2 = \frac{g}{4\pi^{2}}\left(\frac{1}{\Gamma_{1,\,0}}\frac{\mathrm{d}\ln P}{\mathrm{d}r} - \frac{\mathrm{d}\ln \rho}{\mathrm{d}r}\right)\,.
\end{equation}

In these expressions, $g$ is the local gravitational acceleration, $P$ the pressure, $\rho$ the density, $c_{\mathrm{s}}$ the local sound speed, $\Gamma_{1,\,0}$ the first adiabatic exponent, and $r$ the radial coordinate. It is instructive to rewrite the expression for $\Tilde{N}$ in an approximate form applicable for fully ionised ideal gasses:
\begin{equation}
    \Tilde{N}^2 \simeq \frac{g^2\rho}{4\pi^{2}P}\left(\nabla_{\mathrm{ad}} - \nabla + \nabla_{\mu} \right)\,,
\end{equation}
with
\begin{equation}
    \nabla = \frac{\mathrm{d}\ln T}{\mathrm{d}\ln P}\,,\quad \nabla_{\mathrm{ad}} = \left(\frac{\mathrm{d}\ln T}{\mathrm{d}\ln P}\right)_{\mathrm{ad}}\,,\quad \nabla_{\mu} = \frac{\mathrm{d}\ln \mu}{\mathrm{d}\ln P}\,,
\end{equation}
$T$ the temperature, and $\mu$ the mean molecular weight.
A g mode with a linear frequency $\nu$ can propagate when $|\nu| < |\Tilde{N}|$ and $|\nu| < \Tilde{S}_{\ell}$. Outside this g-mode cavity, the g modes are evanescent and decay exponentially. For p modes, the mode cavity is determined by the conditions $|\nu| > |\Tilde{N}|$ and $|\nu| > \Tilde{S}_{\ell}$.

In stars where the p- and g-mode cavities overlap in terms of frequencies, such as in evolved massive MS \citep{unno_nonradial_1989} or red giant stars \citep{cunha_structural_2015}, oscillation modes can have a dual p-g character. These p-g mixed modes arise when modes from the p- and g-mode cavity tunnel through the evanescent zone and interact. Since p-g mixed modes have both p- and g-mode characters, we define $n_{\mathrm{g}}$ as the number of nodes in the g character regime and $n_{\mathrm{p}}$ as the number of nodes in the p character regime\footnote{The distinction between p and g nodes is based on whether the spatial derivative of the phase angle $\varphi$, $\mathrm{d}\varphi/\mathrm{d}r$, is positive or negative at the node location, respectively. See \citet{takata_analysis_2006} for more details.}. To uniquely classify p-g mixed modes, we introduce the radial order in the Eckart-Scuflaire-Osaki-Takata scheme, $n_{\mathrm{pg}}$, as \citep{takata_analysis_2006,takata_mode_2012}
\begin{equation}
    n_{\mathrm{pg}} = \begin{cases} 
                        n_{\mathrm{p}} - n_{\mathrm{g}} & \begin{aligned}[t] &\text{for a $\mathrm{g}_{n_{\mathrm{p}} - n_{\mathrm{g}}}$ mode} \end{aligned} \\
                        n_{\mathrm{p}} - n_{\mathrm{g}} + 1 & \begin{aligned}[t] &\text{for a $\mathrm{p}_{n_{\mathrm{p}} - n_{\mathrm{g}}+1}$ mode} \end{aligned}
                      \end{cases}\,.
\end{equation}
Hence, $n_{\mathrm{pg}} = n_{\mathrm{p}}+1$ for pure p modes and $n_{\mathrm{pg}} = -n_{\mathrm{g}}$ for pure g modes.\\

One of the results of the asymptotic theory for nonradial oscillations \citep{tassoul_asymptotic_1980}, which holds for high-order modes ($n \gg 1$), is that the difference in mode periods of g modes with a consecutive number of radial nodes (and the same spherical degree $\ell$), $\Delta P_{n}$, is constant in chemically homogeneous, non-rotating, non-magnetic stars. Here, $\Delta P_{n}$ is defined as 
\begin{equation}\label{eq:PSP}
    \Delta P_{n} = P_{n} - P_{n-1}\,.
\end{equation}
In the expression above, $P_{n}$ is the period of a mode with radial order $n$. For a given degree $\ell$, this constant value is $\Delta P_{n} = \Pi_{\ell}$, with $\Pi_{\ell}$ \citep{aerts_asteroseismology_2010}
\begin{equation}\label{eq:Pil}
    \Pi_{\ell} = \frac{\Pi_{0}}{\sqrt{\ell(\ell+1)}}
\end{equation}
the asymptotic period spacing and 
\begin{equation}\label{eq:buoyancy}
    \Pi_{0} = \pi\left(\int_{r_{\mathrm{i}}}^{r_{\mathrm{o}}}\frac{\Tilde{N}}{r}\mathrm{d}r\right)^{-1}
\end{equation}
the buoyancy travel time \citep{aerts_probing_2021}. The integration boundaries $r_{\mathrm{i}}$ and $r_{\mathrm{o}}$ are the inner and outer turning points of the g-mode cavity, respectively. For a specific mode cavity, the turning points $r_{\mathrm{i}}$ and $r_{\mathrm{o}}$ are defined as the boundaries of the radiative region, that is, the points where $\Tilde{N}$ becomes negative.

We can construct so-called period spacing patterns (PSPs) by plotting observed or theoretically predicted $\Delta P_{n}$ as a function of $n$ or $P_{n}$. Through Eqs.\,(\ref{eq:PSP})--(\ref{eq:buoyancy}), we then get information on the extent of convective regions within observed stars. For example, for MS stars with convective cores and radiative envelopes, the observed $\Delta P_{n}$ is directly related to the extent of the convective core \citep{pedersen_shape_2018,pedersen_internal_2021,michielsen_probing_2019,michielsen_probing_2021,mombarg_asteroseismic_2019,mombarg_constraining_2021,wu_high-precision_2019,wu_asteroseismic_2020}. Even more power of PSPs lies in analysing departures from the uniform value $\Pi_{\ell}$ since they hold information about the star's deep interior. Mode trapping, which can be caused by chemical inhomogeneities \citep{miglio_probing_2008,degroote_deviations_2010} or structural glitches\footnote{These structural glitches refer to sharp features in the BV frequency profile and are related to abrupt changes in the stellar structure. They should not be confused with numerical glitches. See, for example, \citet{aerts_probing_2021} Sect. IV.B for more details.} \citep{cunha_structural_2015}, causes quasi-periodic dips where $\Delta P_{n} < \Pi_{\ell}$ in PSPs. Mode coupling, which is the interaction between modes from different mode cavities or of different nature, is also known to cause dips in PSPs \citep{Mosser2012,cunha_structural_2015,saio_astrophysical_2018, tokuno_asteroseismology_2022, aerts_mode_2023}.

When the effects of rotation are considered, notably the Coriolis acceleration, the stellar oscillation equations become a set of infinitely coupled equations \citep{aerts_asteroseismic_2023}. How to treat this set of coupled equations differs based on how the oscillation mode frequencies relate to the rotation frequency. Modes with $2\pi\nu=\omega > 2\Omega$, with $\omega$ and $\Omega$ the angular mode and rotation frequency, respectively, are called super-inertial modes. We can treat the Coriolis acceleration as a perturbation for super-inertial modes with $\omega \gg 2\Omega$. For sub-inertial g modes, which have $\omega \lesssim 2\Omega$, the Coriolis force acts as an additional restoring force. Such modes are also referred to as gravito-inertial waves (GIWs). For GIWs, the Coriolis force cannot be treated as a perturbation to compute their properties. High-order g modes ($n_{\mathrm{g}} \gg 1$), which typically have low frequencies that obey $\nu \ll \Tilde{N}$ and $\nu \ll \Tilde{S}_{\ell}$, can be treated with the traditional approximation of rotation \citep[TAR,][]{eckart_variation_1960,berthomieu_low-frequency_1978,lee_low-frequency_1987,townsend_asymptotic_2003,mathis_transport_2009}. The TAR is an approximation which neglects the horizontal component of the rotation vector and thus the vertical component of the Coriolis acceleration. This is warranted under the assumption that vertical wave motions are limited by strong chemical and entropy stratification, a condition generally fulfilled in the radiative regions of stars with convective cores. With the TAR, the stellar oscillation equations decouple and can be rewritten as the Laplace tidal equation \citep{laplace_traite_1799}. Theoretical period spacing patterns for gravito-inertial modes under the TAR can be computed as

\begin{equation}\label{eq:psp_tar}
    (\Pi_{s})_{\mathrm{co}} \simeq \frac{\Pi_{0}}{\sqrt{\Lambda_{s,\,l,\,m}}\left(1+\frac{1}{2}\frac{\mathrm{d}\ln\Lambda_{s,\,l,\,m}}{\mathrm{d}\ln s}\right)}\,.
\end{equation}
In the above expression, $(\Pi_{s})_{\mathrm{co}}$ is the period spacing in the co-rotating frame for a mode with spin parameter $s = 2\Omega/\omega$ and $\Lambda_{s,\,l,\,m}$ is the corresponding eigenvalue of the Laplace tidal equation. To go from the co-rotating to the inertial frame in which stars are observed, we use the following relation \citep{bouabid_effects_2013}:
\begin{equation}
    P_{\mathrm{in}} = \frac{P_{\mathrm{co}}}{1+m\frac{P_{\mathrm{co}}}{P_{\mathrm{rot}}}}\,,
\end{equation}
where $P_{\mathrm{in}}$ and $P_{\mathrm{co}}$ are the mode periods in the inertial and co-rotating frame, respectively, and $P_{\mathrm{rot}} = 2\pi/\Omega$ is the rotation period. Since $(\Pi_{s})_{\mathrm{co}}$ in Eq.\,(\ref{eq:psp_tar}) is mode dependent (through the spin parameter $s$), period spacing patterns for GIWs are no longer uniform but have slopes, as demonstrated in \citet{bouabid_effects_2013}. In general, PSPs of prograde modes, which travel with the star's rotation and have $m>0$, have negative slopes in the co-rotating frame. PSPs of modes travelling against the star's rotation or retrograde modes ($m<0$) have positive slopes in the co-rotating frame. Within the framework of the TAR, it has been possible to constrain internal rotation profiles for hundreds of stars with convective cores from measured PSPs of gravito-inertial waves \citep{VanReeth2015a,VanReeth2015b,van_reeth_interior_2016,ouazzani_new_2017,van_reeth_sensitivity_2018,li_period_2019,li_gravity-mode_2020, van_reeth_near-core_2022}. 

\section{Methods}\label{sec:methods}
We computed the asteroseismic properties of merger products and genuine single stars by solving the stellar oscillation equations, which require an equilibrium stellar structure model as input. These equilibrium stellar structure profiles were taken from 1D stellar evolution models. We used non-rotating equilibrium models to predict the asteroseismic properties with the inclusion of rotation at the level of the stellar oscillation equations. Ignoring rotation in the equilibrium models disregards the theory of rotationally induced mixing, but we mimic its effects by means of simpler approximations for the internal mixing profiles \citep{pedersen_internal_2021}. Following 
\citet{Henneco2021} we also ignored the centrifugal deformation of the equilibrium model, because it results in negligible frequency shifts. \citet{Henneco2021} furthermore showed that for up to $70\%$ of critical rotation, the inclusion of the centrifugal deformation at the level of the asymptotic pulsation mode predictions and the level of the equilibrium model combined lead to fractional frequency shifts well below $1\%$. Therefore, the effect of the centrifugal deformation may be neglected for initial asteroseismic modelling attempts for rotation rates up to $70\%$ of the critical rotation rate. The procedure of including rotation only at the level of the stellar oscillation equations and ignoring the centrifugal deformation of the equilibrium model is common practice \citep{aerts_probing_2021} and is further elaborated in \citet{aerts_asteroseismic_2023}, to which we refer for details. Section \ref{sec:equilibrium_models} describes the computation of the equilibrium models for the genuine single HG stars and post-MS merger products. We show in Sect.\,\ref{sec:oscillation_equations} how we predicted the asteroseismic properties by solving the oscillation equations using the equilibrium models as input. The input files for the various codes used in this work are available online\footnote{\url{https://zenodo.org/doi/10.5281/zenodo.12087024}}. 

\subsection{Stellar model computations with \texttt{MESA}}\label{sec:equilibrium_models}
\subsubsection{Adopted stellar physics}\label{sec:aptopted_physics}
We used the stellar structure and evolution code \texttt{MESA} \citep[r12778;][]{paxton_modules_2011,paxton_modules_2013,paxton_modules_2015,paxton_modules_2018,paxton_modules_2019} to compute non-rotating single-star evolution models at solar metallicity ($Y = 0.2703$ and $Z=0.0142$, \citealt{asplund_chemical_2009}). We used a combination of the OPAL \citep{iglesias_radiative_1993,iglesias_updated_1996} and \citet{ferguson_low-temperature_2005} opacity tables suitable for the chemical mixture of \citet{asplund_chemical_2009}. We did not enable \texttt{MESA}'s hydrodynamic solver, that is, all models are hydrostatic. We used the \texttt{approx21} nuclear network. Each model was initialised at its Zero-Age Main Sequence (ZAMS) and evolved until core-helium exhaustion (i.e. when the central mass fraction of helium is below $10^{-6}$).\\

The Ledoux criterion was employed to determine which regions of the stellar model were convective and the mixing length theory \citep[MLT,][]{bohm-vitense_uber_1958,cox_principles_1968} for the treatment of convective mixing, with a mixing length parameter of $\alpha_{\mathrm{mlt}} = 2.0$ as the best estimate from asteroseismology of stars on the MS with a convective core \citep[e.g.][]{fritzewski_age_2024}. Semi-convective mixing is included with an efficiency of $\alpha_{\mathrm{sc}}=10.0$ \citep{schootemeijer_constraining_2019}. For thermohaline mixing, we used $\alpha_{\mathrm{th}}= 1.0$ \citep{marchant_role_2021}. We added a constant envelope mixing of $\log(D_{\mathrm{mix}}/\mathrm{cm}^2\mathrm{s}^{-1}) = 3$ to smooth out small step-like features in the chemical composition left behind by the receding convective cores and shrinking convective shells. This value used for $D_{\mathrm{mix}}$ is typical for what is deduced from asteroseismology of B stars \citep{pedersen_internal_2021}.

Convective boundary mixing (CBM) was handled through the overshooting scheme. For hydrogen-burning convective cores, we used the step overshooting scheme in \texttt{MESA} to extend the convective region by $0.20H_{P}$ \citep{martinet_convective_2021} beyond the boundary set by the Ledoux criterion. Here, $H_{P}$ is the pressure scale height. We used exponential overshooting \citep{Herwig2000} for helium-burning cores, with $f_{\mathrm{ov}} = 0.015$. The magnitude of overshooting above helium-burning cores is not constrained adequately and yet can severely influence the final fate of stars \citep{Temaj2024,Brinkman2024}. We chose a moderately high value in a range consistent with observations of intermediate- \citep{constantino_treatment_2016} and low-mass \citep{bossini_kepler_2017} stars. This value of $f_{\mathrm{ov}}$ was chosen to avoid the occurrence of breathing pulses, which are instabilities of the convective helium-burning core that occur in models with lower values of overshooting. Whether these breathing pulses are numerical artefacts or physical instabilities is unclear, yet recent evidence supports the former \citep{ostrowski_evolutionary_2021}. Above and below convective shells and below convective envelopes, we used exponential overshooting with $f_{\mathrm{ov}} = 0.005$, which is consistent with values typically inferred for the Sun \citep{angelou_convective_2020}. 

We used the same setup as in \citet{henneco_contact_2024} to compute the wind mass-loss rate. For hot stars ($T_{\mathrm{surf}} \geq 11\,\mathrm{kK}$, with $T_{\mathrm{surf}}$ the temperature of the outermost cell), we used the \citet{Vink2000} wind mass-loss prescription with a scaling factor of 1.0. The cool wind regime ($T_{\mathrm{surf}} \leq 10\,\mathrm{kK}$) was divided based on whether the stars evolve into giants or supergiants. The cut is made at $\log(\mathscr{L}/\mathscr{L}_{\odot}) = 3.15$, with $\mathscr{L}$ defined as \citep{langer_spectroscopic_2014}

\begin{align}
    \mathscr{L} = \frac{1}{4\pi\sigma G}\frac{L_{\star}}{M_{\star}}\,.
\end{align}
Here, $G$ is the gravitational constant, $\sigma$ the Stefan-Boltzmann constant, and $L_{\star}$ and $M_{\star}$ the star's luminosity and mass, respectively. We evaluated $\mathscr{L}$ when $T_{\mathrm{surf}} < 11\,\mathrm{kK}$ for the first time during the star's evolution. The division between giant and supergiant wind regimes at $\log(\mathscr{L}/\mathscr{L}_{\odot}) = 3.15$ corresponds roughly to a division at $10\,\mathrm{M}_{\odot}$. We used linear interpolation to compute the wind mass-loss rate between the hot and cool wind regimes.

\subsubsection{Merger products via fast accretion}\label{sec:fast_accretion}
This work focuses on massive stars produced in early Case B (Case Be) mergers. These stellar mergers occur when the initially more massive or primary star is in the HG and does not yet have a (deep) convective envelope \citep{henneco_contact_2024}. To produce a merger product in \texttt{MESA}, we evolved a single star, using the assumptions described above, until it reached the HG. When the star reached a point in the HG at which its effective temperature $T_{\mathrm{eff}}$ is cooler than roughly its lowest MS value, we invoked the merger procedure.
The merger procedure consisted of accreting a specified mass $\Delta M$ onto the single star with initial mass $M_{\mathrm{i}}$ on a timescale of $0.1\tau_{\mathrm{KH}}$, with $\tau_{\mathrm{KH}}$ the star's current global thermal or Kelvin-Helmholtz timescale, defined as \citep{kippenhahn_stellar_2013}
\begin{equation}\label{eq:thermal_timescale}
    \tau_{\mathrm{KH}} = \frac{GM_{\star}^{2}}{2R_{\star}L_{\star}} \approx 1.5\times10^{7}\left(\frac{M_{\star}}{\mathrm{M}_{\odot}}\right)^{2}\frac{\mathrm{R}_{\odot}}{R_{\star}}\frac{\mathrm{L}_{\odot}}{L_{\star}}\mathrm{yr} \,.
\end{equation}  
Here, $R_{\star}$ is the stellar radius. This procedure is similar to what is used in \citet{Justham2014}, \citet{rui_asteroseismic_2021}, \citet{Deheuvels2022}, and \citet{schneider_pre-supernova_2024} to mimic the result of stellar mergers.
\citet{Justham2014} assumed accretion happens on a timescale ${\lesssim}10^{4}\,$ yr,
\citet{rui_asteroseismic_2021} used a fixed accretion rate of $10^{-5}\,\mathrm{M}_{\odot}\,\mathrm{yr}^{-1}$, while 
\citet{schneider_pre-supernova_2024} assumed accretion to happen on the star's thermal timescale, that is, at an accretion rate $\dot{M}_{\mathrm{acc}} = M_{\star}/\tau_{\mathrm{KH}}$.  Although the technical setup is essentially the same, \citet{Deheuvels2022} studied the effect of accretion in a binary system rather than for stellar mergers, which occur on shorter timescales. \citet{schneider_pre-supernova_2024} used their setup to study both merger and accretion products.

During the fast accretion phase, an extended convection region develops in the star's envelope. After this phase, we mix the outer envelope, that is, the region from the top of the convective hydrogen-burning shell to the surface with $\log(D_{\mathrm{mix}}/\mathrm{cm}^2\mathrm{s}^{-1}) = 12$ for a time $0.01\tau_{\mathrm{KH}}$. We did this to smooth out the abrupt changes in the chemical composition profile left behind by the extended convection zone. Afterwards, we evolved the merger models until the end of core helium exhaustion.

\subsubsection{Limitations of the fast accretion method}\label{sec:limitations}
Stellar mergers are complex phenomena that include a wealth of physical processes. 3D merger simulations, such as those in \citet{Lombardi2002}, \citet{Ivanova2002}, \citet{Glebbeek2013}, and \citet{schneider_stellar_2019} currently are our best windows into the merging process and the products that result from it, but they are computationally expensive. As a result, these 3D simulations are limited to only a handful of initial binary parameters. The fast accretion method provides a quick and flexible zeroth-order approximation for merger product structures. We now list some of its main limitations.

First, before stars merge, they evolve through a contact phase, preceded by a mass-transfer phase (this is true if we consider a merger driven by binary evolution channels and not through dynamical interactions). During both phases, the structure can be altered significantly \citep{henneco_contact_2024}. With the fast accretion method, we assume that the star onto which the companion is accreted is unaltered by any previous mass-transfer and contact phases.

Second, fast accretion does not account for the chemical composition of the merger product because the chemical composition of the accreted material is taken to be the same as the surface chemical composition of the accretor. With this assumption, we primarily underestimate the amount of helium in the envelope of the merger product. Moreover, it does not account for the mixing of stellar material from the two components during the merger phase. As a result, neither the internal chemical structure nor the surface chemical abundances can be reproduced correctly by the fast accretion method. We stress that we use the fast accretion method to create effective merger product structures of the kind in which the less evolved secondary star is mixed in with the more evolved primary star's envelope. Generally, mass is expected to be lost from both components during the merger phase (see, e.g. \citealt{schneider_stellar_2019}). Therefore, $\Delta M$ is the mass effectively added to the primary's envelope. The added mass fraction $f_{\mathrm{add}} = \Delta M/M_{\mathrm{i}}$ should not be confused with the mass ratio of the merger product's progenitor binary system. From smoothed particle hydrodynamic (SPH) simulations of mergers \citep[e.g.,][]{Lombardi2002,Gaburov2008,Glebbeek2013}, we know that if the H-rich core of the secondary star has lower entropy than that of the primary star, it can sink to the centre of the merger product. In such cases, the merger product rejuvenates and becomes a MS star \citep{Glebbeek2013}. One needs different merging schemes to create these kinds of merger products, such as entropy sorting \citep{Gaburov2008b}. While such merger products warrant their own investigations (Henneco et al. in prep.), our work focuses solely on long-lived B-type or BSG merger products. 

Third, the fast accretion method does not reproduce the strong surface magnetic fields expected from both 3D magnetohydrodynamics simulations \citep{schneider_stellar_2019} and the recent observation of a massive magnetic star that shows strong signs of being formed in a merger \citep{Frost2024}. Although internal magnetic fields are less well constrained than surface magnetic fields \citep{Donati2009}, it cannot be excluded that they also result from binary mergers. As shown by, for example,  \citet{Prat2019}, \citet{VanBeeck2020}, \citet{Dhouib2022}, and \citet{Rui2024}, internal magnetic fields can significantly influence the frequencies of g modes.

\subsection{Stellar oscillation calculations with \texttt{GYRE}}\label{sec:oscillation_equations}
For the computation of the stellar oscillations, we used the stellar pulsation code \texttt{GYRE} \citep[v7.0;][]{townsend_gyre_2013,townsend_angular_2018} with the equilibrium models produced with \texttt{MESA} as input. We used the \texttt{MAGNUS\_GL6} solver and the boundary conditions from \citet{unno_nonradial_1989} to compute adiabatic\footnote{It is appropriate to use the adiabatic approximation for the computation of oscillation modes in B-type stars \citep{Aerts2018b}.} oscillations for $(\ell,\,m) = (1,\,0)$ and $(\ell,\,m) = (2,\,0)$ modes (no rotation). For the computation of oscillation modes with rotation in the inertial frame, we used the TAR in \texttt{GYRE} (see Sect.\,\ref{sec:diagnostics}). Even though we only consider slow to moderate rotation in this work, the use of the TAR is required (as opposed to treating the Coriolis acceleration as a perturbation), which we demonstrate in Appendix \ref{app:pert_vs_tar}. Core- and envelope rotation rates inferred from asteroseismology show that low- and intermediate-mass stars have nearly uniform radial rotation profiles during their MS and HG or subgiant evolution \citep{Aerts2019}. We currently have internal rotation profiles inferred from asteroseismology for only a handful of high-mass stars, which show core-to-envelope rotation rate ratios between 1 and 5 
without proper error estimation \citep{Burssens2023,aerts_asteroseismic_2023}.  Considering this, we assumed a uniform (solid-body) rotation profile for the \texttt{GYRE} computations. We computed $(\ell,\,m) = (1,\,0)$, $(1,\,\pm 1)$, $(2,\,0)$, $(2,\,\pm 1)$, and $(2,\,\pm 2)$ modes with rotation.

\section{Results}\label{sec:results}

\subsection{Detailed comparison between Case Be merger products and genuine single HG stars}\label{sec:delailed_comparison}

\begin{figure*}[h!]
\centering
  \includegraphics[width=18cm]{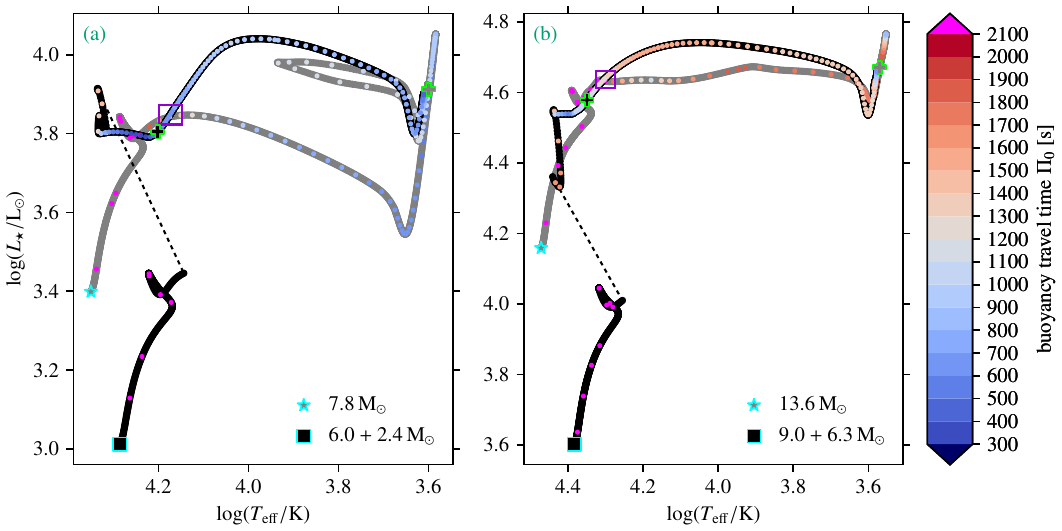}
     \caption{HRD with evolutionary tracks for a merger product of $M_{\star} = 6.0+2.4\,\msun$ and a genuine single star an initial mass of $M_{\star} = 7.8\,\msun$ (Panel a), and a merger product of $M_{\star} = 9.0+6.3\,\msun$ and a genuine single star with an initial mass of $M_{\star} = 13.6\,\msun$ (Panel b). The merger and genuine single-star tracks are shown in black and grey, respectively. To avoid cluttering, the track of the merger model is not shown during the fast accretion phase. A dashed black line replaces it. Each evolutionary track is connected to its label through a symbol on the HRD at the ZAMS. The green plus symbols indicate the position in the HRD after $10^6$ years have passed since the merging or TAMS for the merger product and single star, respectively. The colours on the tracks are related to the buoyancy travel time $\Pi_{0}$. The violet box indicates the position at which we compare the asteroseismic properties of the merger product and genuine single star in Sect.\,\ref{sec:compare1} and \ref{sec:compare2}.}
     \label{fig:hrd_overlap_combo}
\end{figure*}

We now compare the predicted asteroseismic properties of merger products formed through the fast accretion method described in Sect.\,\ref{sec:fast_accretion}, and genuine single stars in the HG. We focus on two merger products resulting from early Case-B mergers: an $M_{\star} = 8.4\,\msun$ product of a star with $M_{\mathrm{i}} = 6.0\,\mathrm{M}_{\odot}$ and $\Delta M = 2.4\,\mathrm{M}_{\odot}$ (added mass fraction $f_{\mathrm{add}} = 0.4$) and an $M_{\star} = 15.3\,\msun$ product of star with $M_{\mathrm{i}} = 9.0\,\mathrm{M}_{\odot}$ and $\Delta M = 6.3\,\mathrm{M}_{\odot}$ ($f_{\mathrm{add}} = 0.7$). We compare both merger products with appropriate genuine single-star counterparts. To find these genuine single star counterparts, we compared the merger products' HRD tracks with a range of genuine single star tracks of different masses and selected those which cross in the blue region of the HRD ($\log T_{\mathrm{eff}}/\mathrm{K} \gtrsim 4.0$). Hence, we compare the $6.0+2.4\,\msun$ and $9.0+6.3\,\msun$ merger products with $7.8\,\msun$ and $13.6\,\msun$ genuine single stars, respectively.

The evolutionary tracks of the merger products and genuine single stars are shown in Fig.\,\ref{fig:hrd_overlap_combo}. Each track has been colour-coded with the value of the buoyancy travel time $\Pi_{0}$, defined in Eq.\,(\ref{eq:buoyancy}). Each HRD track is also marked with a plus symbol, which indicates the position of the star 1\,Myr ($10^{6}\,\mathrm{yr}$) after the terminal-age main sequence (TAMS) for genuine single stars, and 1\,Myr after the merger event for merger products. Given that the plus symbol is located in the blue region of the HRD for the merger products, it is clear that they spend at least 1\ Myr there and are, hence, likely to be observed as B-type stars or BSGs. Genuine single stars have already moved towards the red region of the HRD within 1\,Myr and are, therefore, less likely to be observed as B-type stars or BSGs. The merger products' luminosities increase during their evolution on the HG (around $\log T_{\mathrm{eff}}/\mathrm{K} = 4.2$ for the $6.0+2.4\,\msun$ product and $\log T_{\mathrm{eff}}/\mathrm{K} = 4.4$ for the $9.0+6.3\,\msun$ product), which is caused by the onset of core He burning. We define the onset of core He burning as when the central He mass fraction falls below $99\%$ of its value at the TAMS. For the $13.6\,\msun$ genuine single star, core He ignition also occurs already on the HG around $\log T_{\mathrm{eff}}/\mathrm{K} = 4.1$. Our genuine single stars with masses $\gtrsim 11.4\,\msun$ ignite helium in their cores on the HG.

\subsubsection{Comparison between a $6.0+2.4\,\msun$ merger product and a $7.8\,\msun$ genuine single star}\label{sec:compare1}

\begin{figure*}
\centering
  \includegraphics[width=18cm]{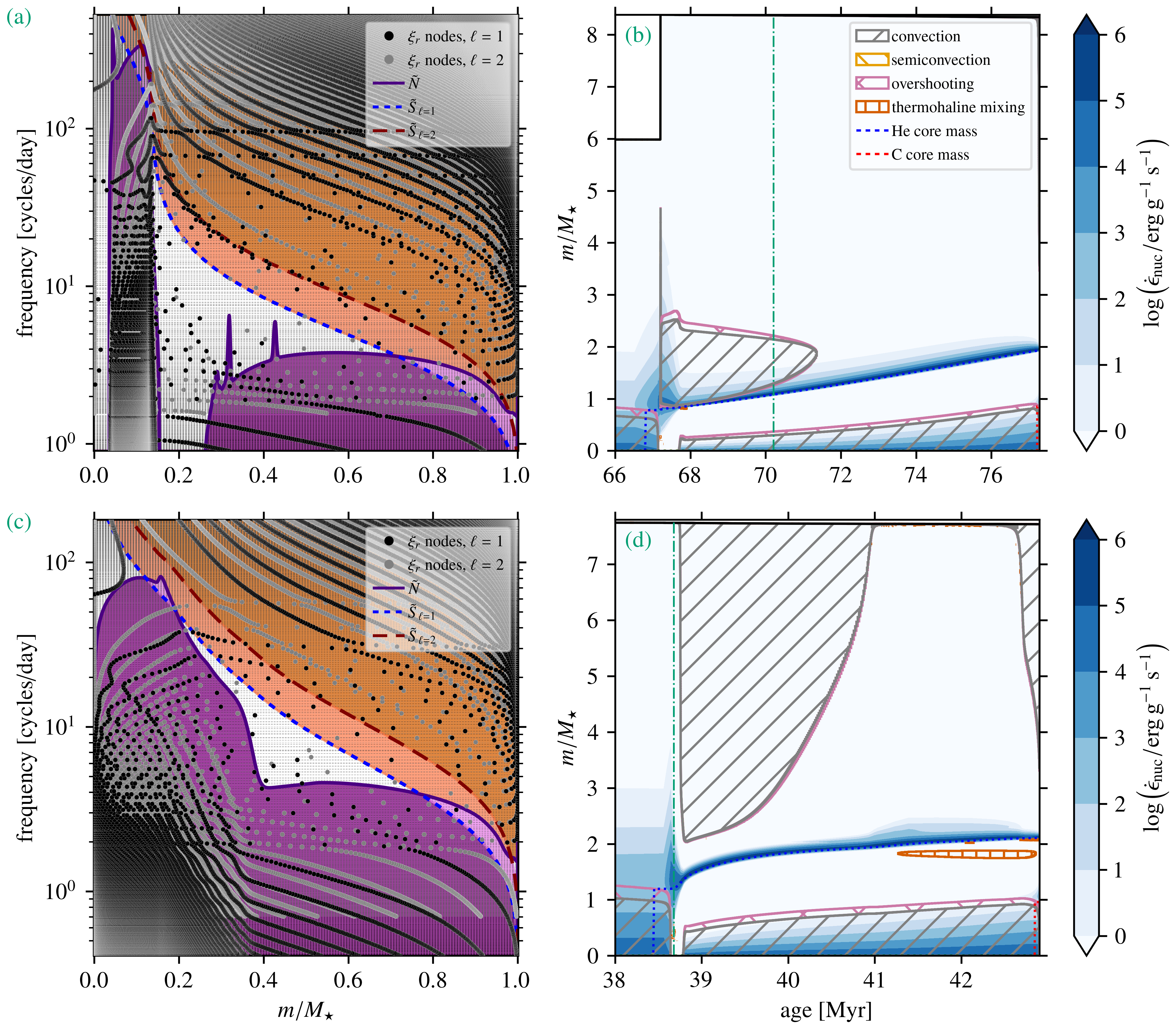}
     \caption{Panel (a) and (c) show propagation diagrams for $\ell = 1,\,2$ and $m=0$ modes without rotation for the merger product with $M_{\star} = 6.0+2.4\,\msun$ and genuine single star with $M_{\star} = 7.8\,\msun$, respectively. The black (grey) dots represent the radial nodes of the oscillation modes, or more specifically, the locations where the radial wave displacement $\xi_{r}(r) = 0$ for the $\ell = 1$ ($\ell = 2$) modes. The purple regions on this diagram show the g-mode cavity or cavities, while the orange region shows the p-mode cavity. Panel (b) and (d) show Kippenhahn diagrams for the merger product with $M_{\star} = 6.0+2.4\,\msun$ and genuine single star with $M_{\star} = 7.8\,\msun$, respectively. The green dash-dotted lines indicate the models for which the respective propagation diagrams in Panel (a) and (c) are shown, which is when their HRD tracks overlap, indicated by the violet box in Fig.\,\ref{fig:hrd_overlap_combo}a. Both Kippenhahn diagrams show the evolution of the models up to core-helium exhaustion.}
     \label{fig:propkip_compare1}
\end{figure*}

\begin{figure*}
    \sidecaption
    \includegraphics[width=12cm]{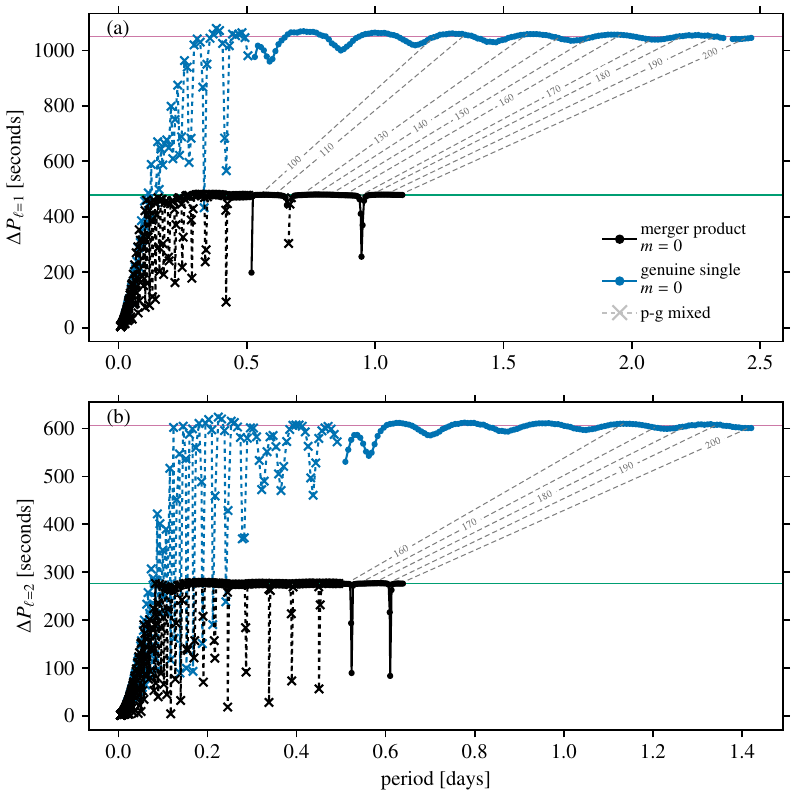}
    \caption{Period spacing patterns for the $6.0+2.4\,\mathrm{M}_{\odot}$ merger product (black) and the $7.8\,\mathrm{M}_{\odot}$ genuine HG star (blue) without rotation at the time of comparison (violet box in Fig.\,\ref{fig:hrd_overlap_combo}). The dashed grey lines connect pure g modes with the same radial order $n_{\mathrm{g}}$. The black and grey dashed lines with cross symbols indicate p-g mixed modes. These p-g mixed modes have at least one node in the radial direction, that is, $n_{\mathrm{p}} > 0$. Panel (a) and (b) show the period spacing patterns for $(\ell,\,m) = (1,\,0)$ and $(2,\,0)$ modes, respectively. The green and purple horizontal lines indicate the $\Pi_{\ell}$ values for the merger product and genuine single star, respectively.}
    \label{fig:psps_compare1}
\end{figure*}

We find a clear difference between the respective $\Pi_{0}$ values of the merger product and genuine single star during the time their evolutionary tracks cross in the HRD, indicated by the violet box in Fig.\,\ref{fig:hrd_overlap_combo}a. The merger product has $\Pi_{0}\ = 678\,\mathrm{s}$ and for the genuine single star we find $\Pi_{0} = 1485\,\mathrm{s}$. This difference comes from the fact that the g-mode cavities, which determine the value of $\Pi_{0}$ (Eq.\,\ref{eq:buoyancy}), have significantly different shapes. From the propagation\footnote{Alternative versions of the propagation diagrams as a function of the relative radial coordinate $r/R_{\star}$ can be found in Appendix \ref{app:prop_radial}.} and Kippenhahn diagrams in Fig.\,\ref{fig:propkip_compare1}, we see that the merger product has two g-mode cavities, one near the convective He-burning core (inner cavity) and one further out (outer cavity). Because of the mass accretion onto the primary star during the merger, it has a higher envelope mass and, consequently, a higher temperature at the base of the envelope. The higher temperature leads to a larger H-shell burning luminosity. This then drives a convective zone in and above the H-burning region. This convective region is responsible for separating the g-mode cavity into two parts. The genuine single star is fully radiative and hence has a single g-mode cavity (Figs.\,\ref{fig:propkip_compare1}c and \ref{fig:propkip_compare1}d). From Fig.\,\ref{fig:propkip_compare1}a, we also see that the pure g modes are mostly confined to the inner cavity and only have a few radial nodes in the outer cavity. To correctly apply Eq.\,(\ref{eq:buoyancy}), which is derived within the asymptotic theory ($n \gg 1$), we only integrate over the inner cavity. Integrating over both cavities results in $\Pi_{0}$ values that are lower by up to $10\,\mathrm{s}$, which is on the order of the period precision for time series from space missions such as \textit{Kepler} and the TESS continuous viewing zone \citep{VanReeth2015a,pedersen_internal_2021,garcia_internal_2022,garcia_detection_2022}. Even though the merger product has a smaller mode cavity than the genuine single star, the BV frequency reaches higher values at low radial coordinates (see Figs.\,\ref{fig:prop_radius} and \ref{fig:prop_radius_zoom} in Appendix \ref{app:prop_radial}), which results in a higher value of the integral in the denominator of Eq.\,(\ref{eq:buoyancy}), and hence a smaller value of $\Pi_{0}$ compared to the genuine single star. 

The difference in terms of asteroseismic properties between the merger product and genuine single star can be further appreciated from the comparison between their PSPs for $(\ell,\,m) = (1,\,0)$ and $(2,\,0)$ modes without rotation and with radial orders $n_{\mathrm{pg}}$ between -1 and roughly -200 (Fig.\,\ref{fig:psps_compare1}). We see that for high radial orders (long mode periods $P_{n}$), the mean values of the PSPs differ significantly, as expected from our earlier estimation based on $\Pi_{0}$. Next to the difference in mean PSP values, we see that modes with the same number of radial nodes have different mode periods in the two models. The modes of the merger product have shorter mode periods for the same number of nodes compared to the modes of the genuine single star. This period shift increases with increasing mode period (increasing radial order $n_{\mathrm{pg}}$), both for $(\ell,\,m) = (1,\,0)$ and $(2,\,0)$ modes. Lastly, we see another clear difference between the respective PSPs of the merger product and genuine single star in the form of relatively deep and narrow dips. These dips arise whenever a star has two g-mode cavities. We elaborate further on the nature of these deep dips in Sect.\,\ref{sec:deepdips}.  

\subsubsection{Comparison between a $9.0+6.3\,\msun$ merger and a $13.6\,\msun$ genuine single star}\label{sec:compare2}

\begin{figure*}
\centering
  \includegraphics[width=18cm]{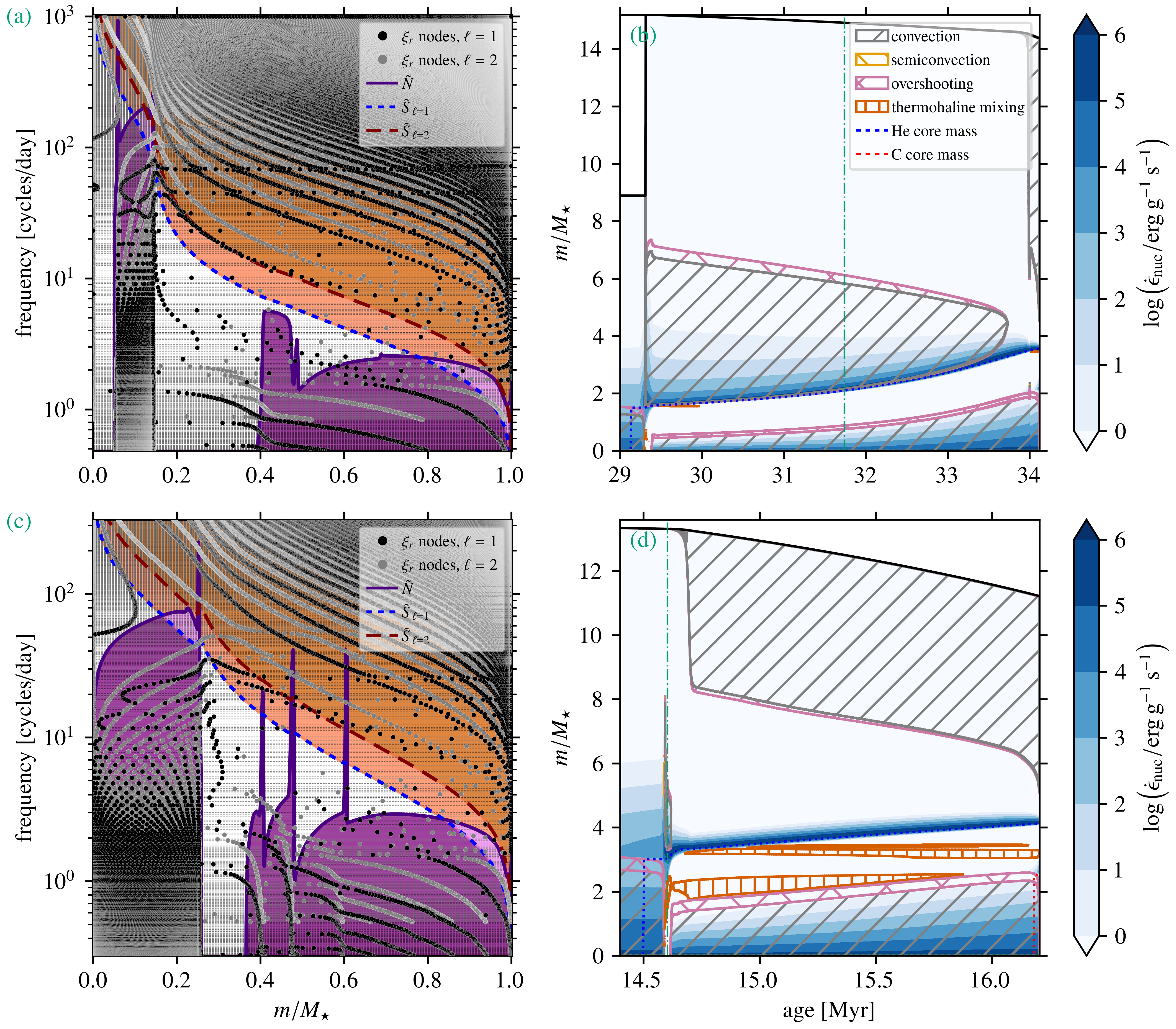}
     \caption{Same as Fig.\,\ref{fig:propkip_compare1}, now for the merger product with $M_{\star} = 9.0+6.3\,\msun$ and genuine single star with $M_{\star} = 13.6\,\msun$.}
     \label{fig:propkip_compare2}
\end{figure*}

\begin{figure*}
    \sidecaption
    \includegraphics[width=12cm]{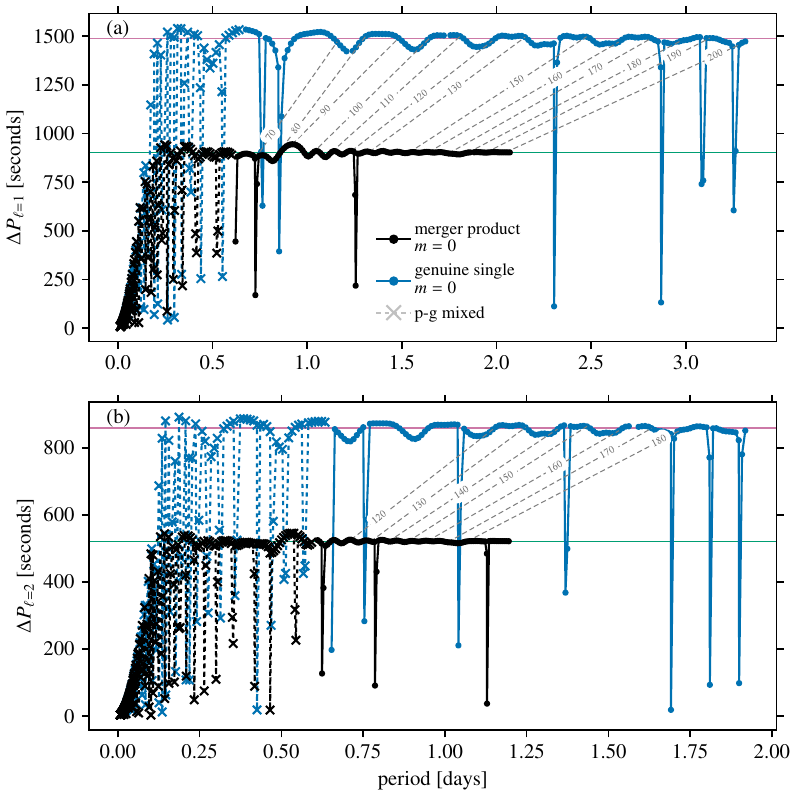}
    \caption{Same as Fig.\ref{fig:psps_compare1}, now for a $9.0+6.3\,\mathrm{M}_{\odot}$ merger product (black) and a $13.6\,\mathrm{M}_{\odot}$ genuine HG star (blue).}
    \label{fig:psps_compare2}
\end{figure*}

Figure \ref{fig:hrd_overlap_combo}b shows the HRD tracks for a $9.0+6.3\,\mathrm{M}_{\odot}$ merger product and a $13.6\,\mathrm{M}_{\odot}$ genuine single star. During their time in the blue side of the HRD, both stars are observable as BSGs ($\log L_{\star}/\mathrm{L}_{\odot} \gtrsim 4.0$\footnote{The lower luminosity limit for the BSGs is not well defined. For this work we use a limit based on observed BSGs.}, \citealt{Urbaneja2017,Bernini-Peron2023}). The $\Pi_{0}$ values for the merger product and genuine single star when their HRD track cross (indicated by the violet box in Fig\,\ref{fig:hrd_overlap_combo}b) are $\Pi_{0} = 1276\,\mathrm{s}$ and $\Pi_{0} = 2104\,\mathrm{s}$, respectively. As expected from \citet{mombarg_asteroseismic_2019} and \citet{pedersen_internal_2021}, these values are higher than for their lower mass analogues described in Sect.\,\ref{sec:compare1}. The absolute difference between the $\Pi_{0}$ values is $828\,$s for this comparison. For the lower-mass counterparts described in Sect.\,\ref{sec:compare1}, the absolute difference between the $\Pi_{0}$ values is $807\,$s. These values are of the same order of magnitude, while the absolute difference is somewhat larger for the more massive merger product and genuine single star. At slightly lower effective temperatures, the values of $\Pi_{0}$ become more comparable than at the effective temperatures the $9.0+6.3\,\msun$ merger product and $13.6\,\msun$ genuine single star models are compared, but they remain distinguishable based on their asymptotic period spacing. The propagation and Kippenhahn diagrams for these models (Fig.\,\ref{fig:propkip_compare2}) show that their structures are more comparable than their lower-mass counterparts. Notably, the $13.6\,\msun$ genuine single star has a convective shell above the H-burning shell, the so-called intermediate convective zones (ICZs). The extent and lifetime of these ICZs are sensitive to the assumptions for convective mixing in the stellar models \citep[e.g.][see also Appendix \ref{app:effect_sc}]{kaiser_relative_2020,sibony_impact_2023}. Because of the ICZ, the $13.6\,\msun$ genuine single star has two g-mode cavities, just like the merger product. The main difference between the two models remains, as for the lower-mass counterparts, the absence of a convective core in the structure of the genuine single star, which has the largest influence on the value of $\Pi_{0}$.

Figure \ref{fig:psps_compare2} shows the PSPs for the $9.0+6.3\,\mathrm{M}_{\odot}$ merger product and the $13.6\,\mathrm{M}_{\odot}$ genuine single star. The PSPs for the merger product look similar to those of its lower-mass counterpart (see Fig.\,\ref{fig:psps_compare1}). We find some key differences when comparing the PSPs of the genuine single star with those of its lower-mass counterparts. First, we see the presence of deep dips with similar morphologies as those in the merger product's PSPs. As mentioned in the previous section, these are related to the existence of an inner and outer g-mode cavity. Second, we see relatively strong quasi-periodic wave-like variability in the $\Delta P_{n}$ values. Such variability is also present for pure g modes in the $7.8\,\msun$ genuine single star (Fig.\,\ref{fig:psps_compare1}), but to a lesser extent. As discussed in Sect.\,\ref{sec:diagnostics}, such wave-like variability in PSPs is caused by mode trapping \citep{pedersen_shape_2018,michielsen_probing_2019,michielsen_probing_2021}, which itself can be caused by sharp features or structural glitches in the $\Tilde{N}(r)-$profile. We see from Fig.\,\ref{fig:propkip_compare2}c that both the inner and outer g-mode cavities of the $13.6\,\msun$ genuine single star have prominent spike features. These features are remnants of an extended, relatively short-lived ($\Delta T \sim 10^{3}\,\mathrm{yr}$), non-uniform (blocky) convection zone that appears after the TAMS and before the development of the ICZ (see Appendix \ref{app:effect_sc}). The non-uniform structure of this convection zone appears at higher and lower spatial and temporal resolutions, and might be the result of an insufficient treatment of convection. We do note that similarly structured convection zones at the onset of the ICZ appear also in the models of \citet{kaiser_relative_2020} for different treatments of convection. We will not discuss the nature of this short-lived, non-uniform convection zone further, but we note that the peaks it introduces in the $\Tilde{N}(r)-$profile influence the oscillation modes. In the outer cavity, where the g-modes only have a handful of nodes, the location of the nodes is influenced by mode trapping caused by the peaks. The strong peaks at the outer edge of the inner g-mode cavity are responsible for the strongest mode trapping and, hence, the quasi-periodic wave-like variability in the PSPs mentioned above. In Appendix \ref{app:disentangle}, we demonstrate how we disentangle the deep dips and quasi-periodic variability in the $13.6\,\msun$ genuine single star's PSP.

\subsection{Deep dips in pure g-mode PSPs}\label{sec:deepdips}

\begin{figure}
    \centering
    \resizebox{\hsize}{!}{\includegraphics{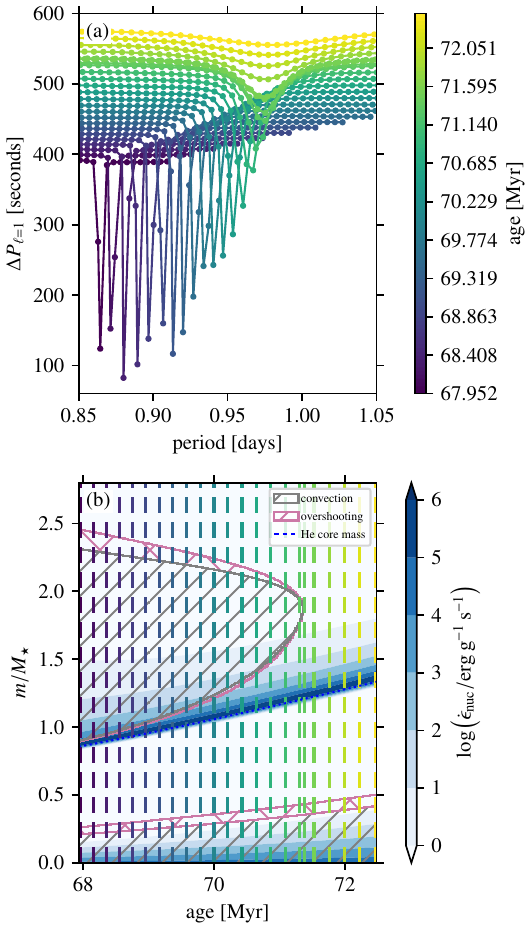}}
    \caption{Temporal evolution of deep PSP dip for the merger product with $M_{\star} = 6.0+2.4\,\msun$. Panel (a) shows the PSP dip for $(\ell,\,m) = (1,\,0$) modes around $0.92$ days. Panel (b) shows a zoom-in of the Kippenhahn diagram from Fig.\,\ref{fig:propkip_compare1}c. The colours of the vertical dashed lines correspond to those in Panel (a).}
    \label{fig:glitches_psp_kipp}
\end{figure}

\begin{figure}
    \centering
    \resizebox{\hsize}{!}{\includegraphics{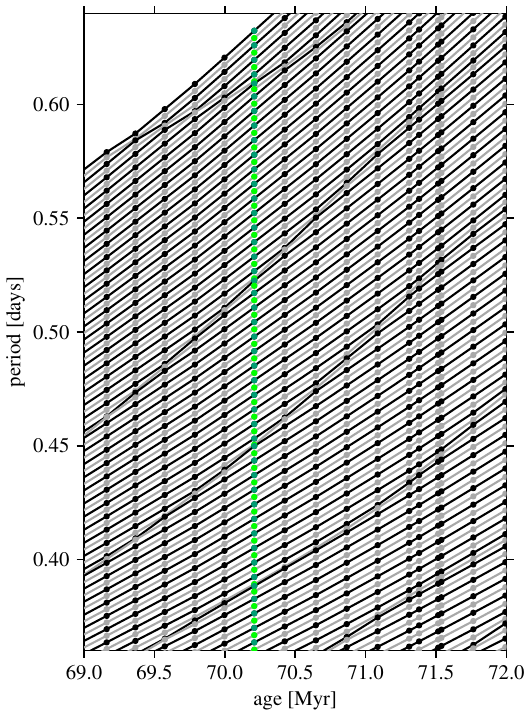}}
    \caption{Zoom-in on a mode-bumping diagram for $(\ell,\,m) = (2,\,0)$ modes of the merger product with $M_{\star} = 6.0+2.4\,\msun$. The black (grey) lines show the evolution of the mode period of g modes with even (odd) radial order $n_{\mathrm{pg}}$. The green (lime) coloured symbols indicate periods of the even (odd) modes shown in Fig.\,\ref{fig:psp_DeltaUnno}.}
    \label{fig:modebumping}
\end{figure}

\begin{figure}
    \centering
    \resizebox{\hsize}{!}{\includegraphics{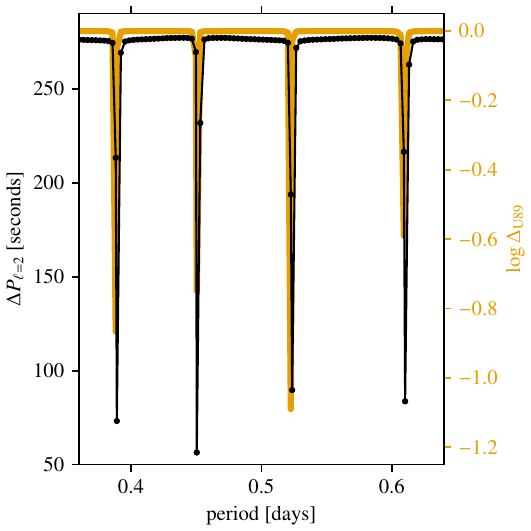}}
    \caption{PSP for $(\ell,\,m) = (2,\,0)$ modes with $n_{\mathrm{pg}} \in [-202;\,-111]$ for the merger product with $M_{\star} = 6.0+2.4\,\msun$ (black line, left axis) and $\Delta_{\mathrm{U89}}$ (orange line, right axis).}
    \label{fig:psp_DeltaUnno}
\end{figure}

From the results in Sect.\,\ref{sec:delailed_comparison}, it has become clear that stellar models with two g-mode cavities, that is, models with either a convective hydrogen-burning shell or ICZ, show deep, narrow dips in their PSPs. In this section, we explore the nature of these deep dips.

First, we look at how these deep dips evolve. From Fig.\,\ref{fig:glitches_psp_kipp}a, we see that the dip location for a specific $(\ell,\,m) = (1,\,0)$ dip moves to longer periods over time. This follows from the deep dip's physical origin. We also see that the width of the dip increases with time and eventually becomes relatively shallow. The transition from a deep, sharp dip to a shallow morphology coincides with the disappearance of the convective zone (Fig.\,\ref{fig:glitches_psp_kipp}b). In other words, the width of the dip is inversely proportional to the width of the evanescent zone between the two g-mode cavities.

Next, we examine the evolution of the g-mode periods with time in Fig.\,\ref{fig:modebumping}. The mode periods (this time shown for $\ell =2 $ modes) increase in time with a quasi-constant slope. For some modes, we observe so-called mode bumping (see, e.g. \citealt{Vanlaer2023}); the period increases faster or slower than the periods of adjacent modes of consecutive radial order, causing the mode periods to be close in value. In such avoided crossings, the mode exchanges energy (couples) with its consecutive mode, which then experiences a faster or slower period increase until it bumps the next mode. This mode bumping sequence continues until the evanescent (convection) zone disappears at around $71.0\text{--}71.5$\,Myr.

Lastly, we show a part of the PSP for $(\ell,\,m) = (2,\,0)$ modes of the $6.0+2.4\,\msun$ merger product in Fig.\,\ref{fig:psp_DeltaUnno}. For each mode in the PSP, we compute the value of $\Delta_{\mathrm{U89}}$, which is the ratio of the kinetic energy of a mode in the inner cavity over its kinetic energy in both cavities \citep{unno_nonradial_1989}
\begin{equation}\label{eq:delta_unno}
    \Delta_{\mathrm{U89}} = \frac{\int^{r_2}_{r_1}4\pi r^{2} \rho\, \left[\xi_{r}(r)^2 + \ell(\ell+1)\xi_{\mathrm{h}}(r)^2\right]\mathrm{d}r}{\int^{R_{\star}}_{0}4\pi r^{2} \rho\, \left[\xi_{r}(r)^2 + \ell(\ell+1)\xi_{\mathrm{h}}(r)^2\right]\mathrm{d}r}\,.
\end{equation}
In this expression, $r_{1}$ and $r_{2}$ are the radii of the inner and outer turning points of the inner g-mode cavity, respectively, and $\xi_{r}(r)$ and $\xi_{\mathrm{h}}(r)$ are the radial and horizontal wave displacement, respectively. For modes that are mostly confined to the inner g-mode cavity, $\Delta_{\mathrm{U89}} \approx 1$, while for modes with a considerable amount of kinetic energy in the outer cavity, $\Delta_{\mathrm{U89}} < 1$. We see from Fig.\,\ref{fig:psp_DeltaUnno} that modes within the deep dips consistently have $\Delta_{\mathrm{U89}} < 1$, meaning that a significant fraction of their kinetic energy sits in the outer g-mode cavity.

Putting the pieces together, we arrive at the nature of the deep dips in the PSPs of models with two g-mode cavities. We find that several inner-cavity g modes tunnel through the evanescent zone, where they interact with outer-cavity g-modes. During this interaction, the coupled modes' periods converge, they exchange energy, and their periods diverge again. The fact that the mode periods converge to the same value causes the deep, narrow dips in the PSPs. The virtual line connecting a sequence of avoided crossings in Fig.\,\ref{fig:modebumping} shows the period evolution of a specific outer-cavity mode. The inversely proportional relation between the dips' width and the evanescent zone's size also follows from this explanation (Fig.\,\ref{fig:glitches_psp_kipp}). Namely, when the evanescent zone is smaller, more inner-cavity modes can tunnel through and couple with outer-cavity modes. This can be appreciated even further when comparing the width of the deep dips for $(\ell,\,m) = (1,\,0)$ and $(\ell,\,m) = (2,\,0)$ modes (Figs.\,\ref{fig:psps_compare1} and \ref{fig:psps_compare2}). Dipole modes ($\ell =1$) have a weaker damping rate than quadrupole ($\ell = 2$) modes \citep[][Chapter 3.4]{aerts_asteroseismology_2010}. Hence, more inner-cavity dipole modes can tunnel through the evanescent zone and interact with the outer-cavity g-modes, leading to wider dips.

\subsection{Effect of added mass fraction $f_{\mathrm{add}}$}\label{sec:q_dep}

\begin{figure}
    \centering
    \resizebox{\hsize}{!}{\includegraphics{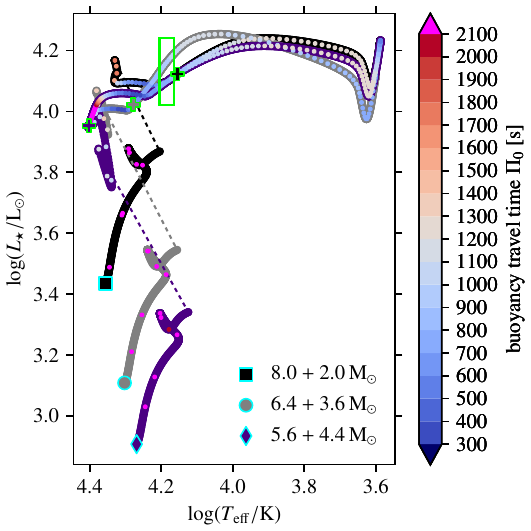}}
    \caption{Same as Fig.\ref{fig:hrd_overlap_combo}, now for $8.0+2.0\,\msun$ ($f_{\mathrm{add}}=0.25$,black square), $6.4+3.6\,\msun$ ($f_{\mathrm{add}}=0.56$, grey circle), and $5.6+4.4\,\msun$ ($f_{\mathrm{add}}=0.79$, indigo diamond) merger products. The lime-coloured rectangle indicates the region in which the three merger products occupy the same region of the HRD and have similar structures.}
    \label{fig:hrd_q_dep}
\end{figure}

\begin{figure}
    \centering
    \resizebox{\hsize}{!}{\includegraphics{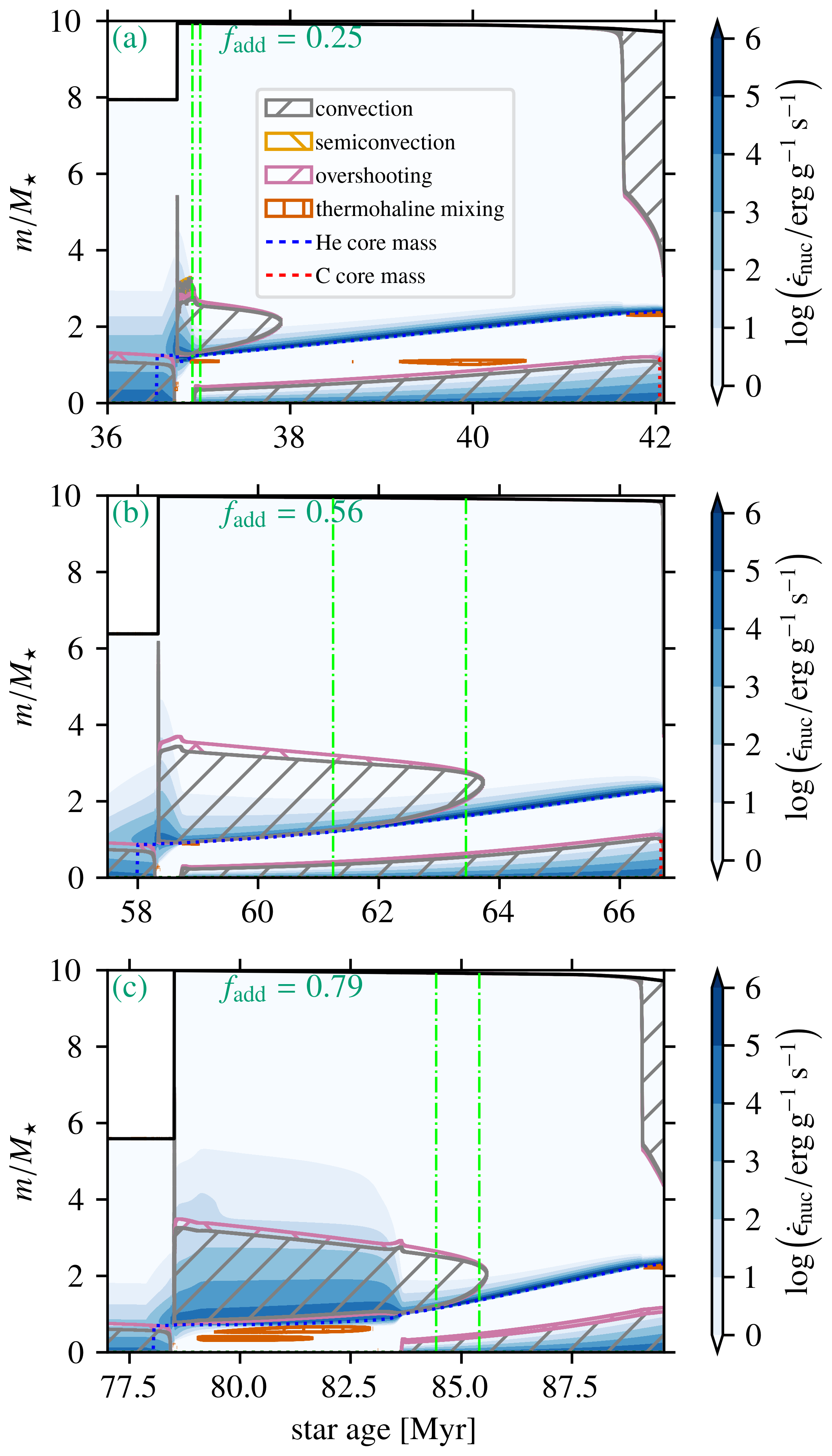}}
    \caption{Kippenhahn diagrams for $8.0+2.0\,\msun$ ($f_{\mathrm{add}}=0.25$, a), $6.4+3.6\,\msun$ ($f_{\mathrm{add}}=0.56$, b), and $5.6+4.4\,\msun$ ($f_{\mathrm{add}}=0.79$, c) merger products. The left (right) lime-coloured vertical line indicates the left (right) bound of the lime-coloured rectangle in Fig.\,\ref{fig:hrd_q_dep}.}
    \label{fig:kipp_q_dep}
\end{figure}

\begin{figure}
    \centering
    \resizebox{\hsize}{!}{\includegraphics{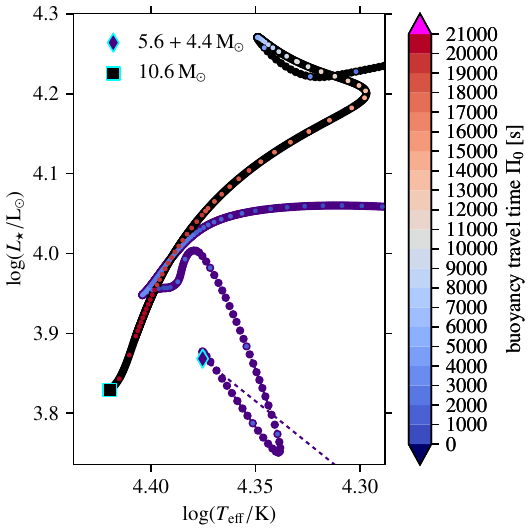}}
    \caption{Zoom-in on the part of the HRD track of the $5.6+4.4\,\msun$ merger product from Fig.\,\ref{fig:hrd_q_dep} that coincides with the MS of a $10.6\,\msun$ genuine singe star. We draw attention to the different colour bar scaling than in Fig.\,\ref{fig:hrd_q_dep}.}
    \label{fig:hrd_q_dep_with_MS}
\end{figure}

As the results in the previous sections have shown, it is, in principle, possible to distinguish merger products from genuine single stars based on their g-mode period spacing patterns. In this section, we briefly explore how the added mass fraction $f_{\mathrm{add}}$ influences our results. We stress again that $f_{\mathrm{add}}$ is a measure of the mass effectively added to the primary star during the merger procedure and should not be confused with the mass ratio of the progenitor binary system (see Sect.\,\ref{sec:limitations}). We compare the HRD tracks of three $10\,\msun$ merger products with varying added mass fractions $f_{\mathrm{add}}$ in Fig.\,\ref{fig:hrd_q_dep}. The following merger products are considered: $8.0+2.0\,\msun$ ($f_{\mathrm{add}} = 0.25$), $6.4+3.6\,\msun$ ($f_{\mathrm{add}} = 0.56$), and $5.6+4.4\,\msun$ ($f_{\mathrm{add}} = 0.79$). Although core He ignition, indicated by the luminosity rise on the HG, occurs at different effective temperatures for the three merger products, they eventually become core-He burning BSGs, which occupy a similar region in the HRD. We indicate this part of the HRD with a lime-coloured rectangle. All merger products have a similar structure during this time: a convective core and a convective H-burning shell. This can be seen in Fig.\,\ref{fig:kipp_q_dep}, in which the vertical lime-coloured lines indicate the age range when the star is located within the lime-coloured rectangle on the HRD in Fig.\,\ref{fig:hrd_q_dep}. The main difference is that models with higher added mass fractions have smaller He-core masses. Despite this, the convective core masses during core He burning are similar because the merger products are all $10\,\msun$ stars that have relaxed to their new structure. This is true for the mass- and $f_{\mathrm{add}}$-ranges considered at the time of the comparison made here, but in general, the final CO core mass depends strongly on $f_{\mathrm{add}}$ \citep{schneider_pre-supernova_2024}. The similar structures of the merger products are reflected in their $\Pi_{0}$ values (Fig.\,\ref{fig:hrd_q_dep}). Closer inspection shows that the values of $\Pi_{0}$ in the overlapping region are $\Pi_{0} = 734\text{--}819\,\mathrm{s}$, $906\text{--}1034\,\mathrm{s}$, and $717\text{--}822\,\mathrm{s}$ for the $8.0+2.0\,\msun$, $6.4+3.6\,\msun$, and $5.6+4.4\,\msun$ merger products, respectively. The $\Pi_{0}$ values for the $6.4+3.6\,\msun$ merger product are considerably higher than for the other merger products, which is related to the fact that by the time it occupies the same region of the HRD as the other merger products, its inner g-mode cavity has shifted outwards because of the growing convective core and shrinking convective H-burning shell. The main cause for the differences in $\Pi_{0}$ is the point in evolution when the models are compared. If we compare the merger products to the left of the lime-coloured rectangle in Fig.\,\ref{fig:hrd_q_dep}, the differences in their $\Pi_{0}$ values are more evident. This is because the merger products ignite helium at different effective temperatures. However, these differences might be method-dependent (see Sect.\,\ref{sec:limitations}), and we opt not to interpret this further than warranted by the nature of our current models. We discuss future steps to improve these comparisons in Sect.\,\ref{sec:discussion}. Lastly, we note from Fig.\,\ref{fig:hrd_q_dep} and Fig.\,\ref{fig:kipp_q_dep}c that the $5.4+4.6\,\msun$ merger product spends a considerable amount of time (between the ages of $79.0$ and $83.5\,$Myr) in the MS region of the HRD. During this time, the merger product has not yet ignited He in its core and has a ${\sim}2\,\msun$ convective H-burning shell. Since such a star would be observed in the same region of the HRD as MS stars, we compare it to a $10.6\,\msun$ genuine single MS star in Fig.\,\ref{fig:hrd_q_dep_with_MS}. During the time the merger product's and MS star's HRD tracks cross, their $\Pi_{0}$ values are $2000\text{--}3000\,$s and $19000\text{--}20000\,$s, respectively. This almost order-of-magnitude difference in $\Pi_{0}$ values follows from the different structures of the stars (see also Eq.\,\ref{eq:buoyancy}): $10.6\,\msun$ MS stars have a convective core and hence a g-mode cavity in the radiative envelope only. The merger product has an inner and outer g-mode cavity, with the bulk of the g modes trapped in and hence sensitive to the inner cavity (see Sect.\,\ref{sec:delailed_comparison})  Therefore, even if a Case-Be merger product is found in the MS region of the HRD, it will be clearly distinguishable from genuine single MS stars based on their respective mean PSP values. This also means that if such merger products contaminate a sample of genuine MS stars, they may influence the inference of the convective core sizes.


\subsection{Comparison for oscillation equations including rotation}
\begin{figure*}
    \sidecaption
    \includegraphics[width=12cm]{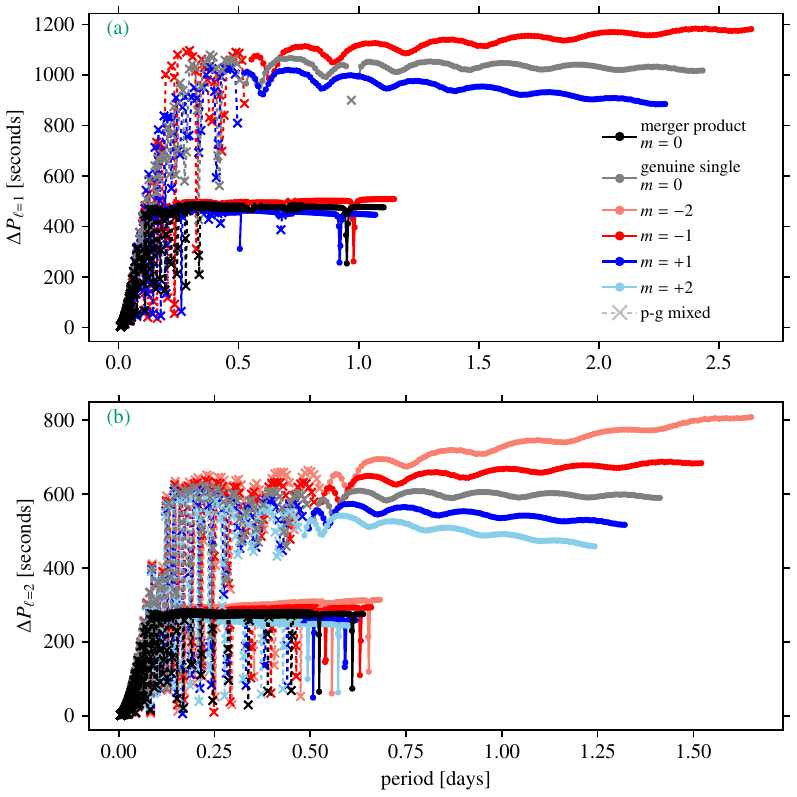}
    \caption{Period spacing patterns for a $6.0+2.4\,\mathrm{M}_{\odot}$ merger product and a $7.8\,\mathrm{M}_{\odot}$ genuine HG star with $\Omega = 0.2\Omega_{\mathrm{c}}$ in the inertial frame. Panel (a) shows the period spacing patterns for $(\ell,\,m) = (1,\,-1)$, $(1,\,0)$, and $(1,\,+1)$ modes, and Panel (b) those for $(\ell,\,m) = (2,\,-2)$, $(2,\,-1)$, $(2,\,0)$, $(2,\,+1)$, and $(2,\,+2)$ modes.}
    \label{fig:psps_compare_Om20}
\end{figure*}

In Fig.\,\ref{fig:psps_compare_Om20} we compare the PSPs of the $6.0+2.4\,\msun$ merger product and $7.8\,\msun$ genuine single star with $\Omega = 0.2\Omega_{\mathrm{c}}$. Here, $\Omega_{\mathrm{c}} = \sqrt{GM_{\star}/R_{\mathrm{eq}}^{3}} \simeq \sqrt{8GM_{\star}/27R^3_{\star}}$ is the Roche critical angular rotation frequency \citep{Maeder2009}. As motivated at the beginning of Sect.\,\ref{sec:diagnostics}, we opted for a relatively slow rotation rate in this work. The angular rotation frequencies for both the merger product and genuine single star correspond to a surface rotation velocity of $v_{\mathrm{surf}} = 38\,\mathrm{km}\,\mathrm{s}^{-1}$. These values for the surface rotation velocity are realistic given the efficient slow-down of stars beyond the TAMS as shown by asteroseismology of single stars \citep{aerts_probing_2021}.

Given the significant difference in the mean PSP values between merger products and genuine single stars, and the relatively low rotation rates, the PSPs remain easily distinguishable with rotation included in the pulsation equations. The appearance of the deep dips in the merger product's PSPs also persists. Their positions shift because of the rotational modulated frequency shifts due to the inclusion of the Coriolis acceleration in the pulsation equations (see \citealt{aerts_asteroseismic_2023} for details).

\subsection{Observability}

We have not yet considered the observability of the modes predicted in this work. 
This depends on many aspects, the most important one being the intrinsic amplitude a mode gets when excited by the physical mechanism responsible for it. Even for 
genuine single intermediate- and high-mass stars, we neither have a complete theory to predict the excitation of the observed gravity modes, nor the intrinsic amplitudes. Indeed, current excitation predictions cannot explain all the nonradial oscillations detected in modern space photometry of such pulsators \citep[e.g.,][for summaries on the shortcomings of the theory revealed by the observations]{HeyAerts2024,Balona2024}. Despite these limitations, especially the lack of knowledge about the modes' intrinsic amplitudes, we perform basic tests of the observability of the oscillation modes predicted in this work.

\subsubsection{Frequency resolution from the observation's time base}

\begin{figure}
    \centering
    \resizebox{\hsize}{!}{\includegraphics[width=9cm]{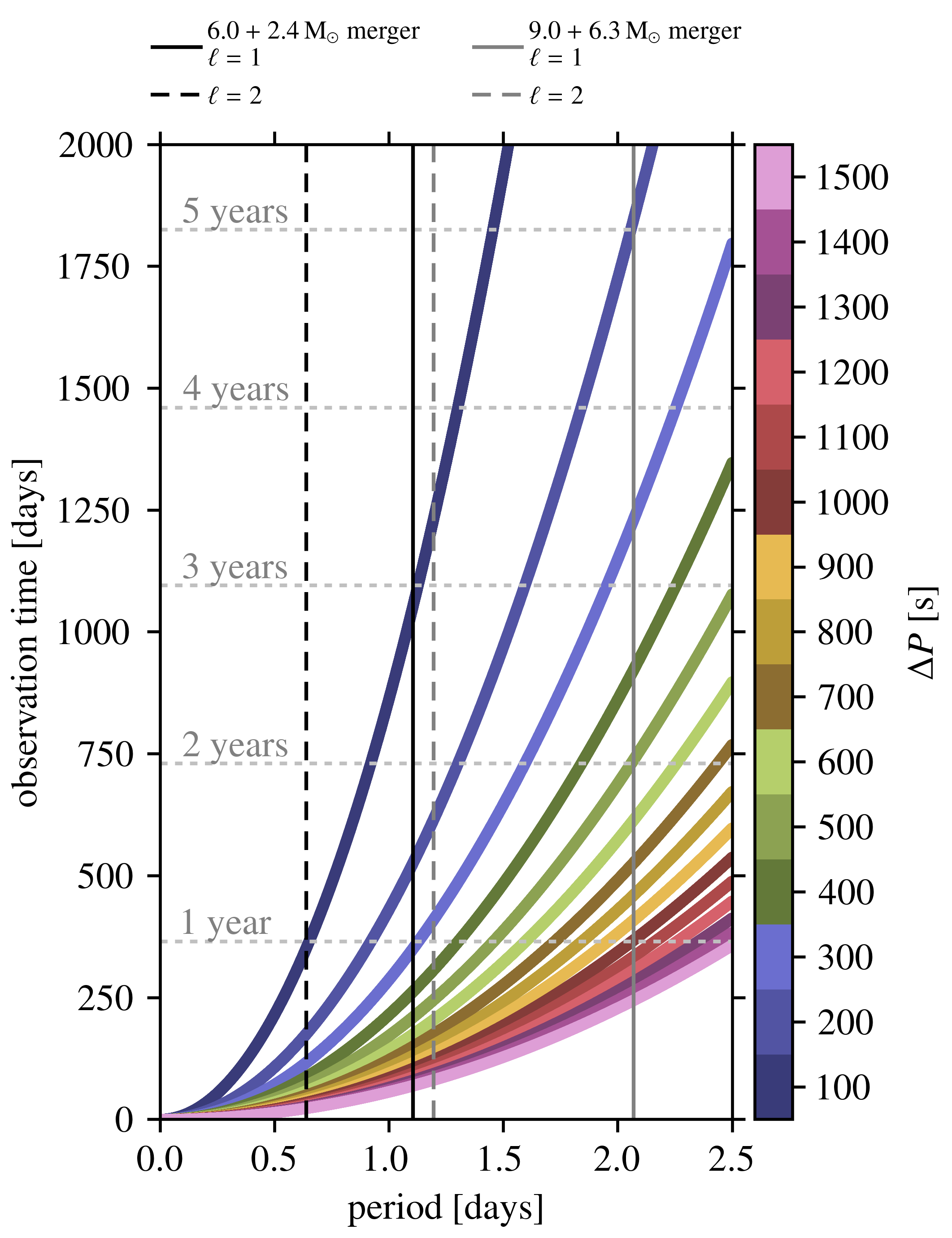}}
    \caption{Observation times required for different period spacings $\Delta P$ in a period range relevant for the period spacing patterns of merger products in our work. The horizontal silver dashed lines indicate the number of years on the y-axis. The vertical solid and dashed lines show the maximum period (i.e. the period for the mode at $n_{\mathrm{pg}} = -200$) for the $(\ell,\,m) = (1,\,0)$ and $(\ell,\,m) = (2,\,0)$ modes, respectively, for the non-rotating $6.0+2.4\,\msun$ (black) and $9.0+6.3\,\msun$ (gray) merger products (see Figs.\,\ref{fig:psps_compare1} and \ref{fig:psps_compare2}).}
    \label{fig:base}
\end{figure}

To estimate the observational baseline required to resolve the period spacings predicted in this work, we start from the fact that the frequency resolution $\delta \nu$ is inversely proportional to the observational baseline $T_{\mathrm{obs}}$
\begin{equation}\label{eq:begin_eq_app}
    \delta \nu = \frac{1}{T_{\mathrm{obs}}}\,.
\end{equation}
Using that the period $P$ is inversely proportional to the frequency $\nu$ and that $\delta \nu = \nu_{2} - \nu_{1}$, we can rewrite Eq.\,(\ref{eq:begin_eq_app}) as
\begin{equation}
    \frac{1}{P_{2}} - \frac{1}{P_{1}} = \frac{1}{T_{\mathrm{obs}}}\,,
\end{equation}
which then gives
\begin{equation}\label{eq:app_final}
    T_{\mathrm{obs}} = \frac{P_{1}P_{2}}{P_{1} - P_{2}} = \frac{P_{1}^{2} - \Delta P P_{1}}{\Delta P}\,,
\end{equation}
where we have used that the period spacing $\Delta P = P_{1} - P_{2}$. We see from Eq.\,(\ref{eq:app_final}) that we require a different observational baseline depending on the period $P_{1}$ and the g-mode period spacing $\Delta P$ we want to observe. We show the results for a range of period spacings and periods relevant for the period spacing patterns of merger products (which have lower PSP values than the genuine single stars) in this work, in Fig.\,\ref{fig:base}. We see that for the $\ell = 1$ modes of the $6.0+2.4\,\msun$ merger product (mean $\Delta P_{n} \approx \Pi_{l} = 479\,\mathrm{s}$), period spacings down to 200\,s can be resolved with observational baselines of two years, and everything down to 100\,s when three years of data are available (for modes with $n_{\mathrm{pg}} \geq -200$). The period spacings for $\ell = 2$ modes (mean $\Delta P_{n} \approx \Pi_{l} = 277\,\mathrm{s}$) should be resolved down to 100\,s with an observational baseline of one year. The mean $\Delta P_{n}$ value for $\ell=1$ modes of the higher-mass $9.0+6.3\,\msun$ merger product is larger than for its lower-mass counterpart (mean $\Delta P_{n} \approx \Pi_{l} = 902\,\mathrm{s}$) and period spacings of $\Delta P \geq 500\,$s between $\ell = 1$ modes with $n_{\mathrm{pg}} \geq -200$ can in principle be resolved with a baseline of two years. For $\ell = 2$ g modes (mean $\Delta P_{n} \approx \Pi_{l} = 521\,\mathrm{s}$), period spacings down to 100\,s can be resolved with an observational baseline of less than four years. Given that inner- and outer-cavity g modes' periods converge when they couple (see Sect.\,\ref{sec:deepdips}), the period spacings in the deep PSP dips can be relatively small. Therefore, resolving all of these dips with reasonable observational baselines might be impossible. However, as discussed in Sect.\,\ref{sec:deepdips}, the deep dips do not consist of a single set of modes. Hence, even when the deep dip's minimum cannot be resolved, the overall deep dip's signature might still be discernible in observational PSPs. We note that we did not consider any form of noise in these estimates of the observational baselines.

\subsubsection{Mode instability for the opacity mechanism}
Stellar oscillation computations in this work have been performed in the adiabatic approximation (see Sect.\,\ref{sec:oscillation_equations}). This approximation does not allow us to predict 
whether modes are excited or damped, that is, which modes are unstable. Even without the adiabatic approximation, \texttt{GYRE} only considers the so-called `heat-engine mechanism' (also known as the opacity or $\kappa$-mechanism) to predict the instability (i.e. the balance between excitation and damping) of modes \citep{aerts_asteroseismology_2010}. While offering a basic understanding for MS pulsators with large amplitudes, this mechanism is known 
to under-predict the number of observed oscillation modes that occur at $\mu$mag level. Many reasons are known to form the basis of the limitations \citep[e.g.\,ignoring radiative levitation in MS B-type pulsators][to mention just one]{Rehm2024}. Other excitation mechanisms include stochastic forcing, convective driving, and non-linear resonant mode excitation as observed in single and close binary pulsators \citep[e.g.,][]{Guo2020,Guo2022,vanbeeck2024}. \citet{DeRidder2023}, \citet{Balona2024}, and \citet{HeyAerts2024} all illustrate that g-mode pulsators form a continuous group of pulsating B-, A-, and F-type stars. Indeed, a significant fraction of these observed pulsators fall outside predicted instability strips based on current mode excitation mechanisms. The situation is even less understood for stars in the HG. In other words, the theory of mode excitation needs to be refined appreciably to explain the observed oscillations in stars in the modern high-cadence, high-precision space photometry for intermediate- and high-mass stars, including mergers.

By solving the oscillation equations for our models with \texttt{GYRE} in its nonadiabatic mode\footnote{The setup for these computational is identical to the one described in Sect.\,\ref{sec:oscillation_equations}, except for that we used the \texttt{MAGNUS\_GL2} solver. This solver is more appropriate for non-adiabatic computations.} and for the current input physics of stellar evolution theory of single and merger stars, we find that none of the oscillation modes in our models are unstable (that is, the imaginary parts of the mode frequency are negative). This is the case in both our merger product and genuine single-star models.

\subsubsection{Wave displacements at stellar surface}

Irrespective of whether the modes treated in this work are predicted to be unstable, the modes' amplitudes throughout the stellar interior and up to the stellar surface can be assessed. 
We provide plots of the wave displacement profiles $\xi_{r}(r)$ and $\xi_{\mathrm{h}}(r)$, and the differential mode inertia profile $\mathrm{d}E/\mathrm{d}r$ for a specific set of oscillation modes for the merger products and genuine single stars described in Sect.\,\ref{sec:compare1} and \ref{sec:compare2}. These plots can be found in Appendix \ref{app:eigen}. The differential mode inertia is the radial derivative of the denominator of Eq.\,(\ref{eq:delta_unno}) (see also Eq.\,3.139 in \citealt{aerts_asteroseismology_2010}). To have a chance to observe these pulsation modes, their wave displacements should not disappear near the surface. 
Although we find that the wave displacements and differential mode inertias are diminished for the pure inner-cavity g modes in stars with two g-mode cavities, there is still a non-negligible mode signal near the surface. The p-g mixed modes, which have shorter mode periods (higher frequencies), couple efficiently, resulting in even larger displacements and differential mode inertias. We stress that this does not immediately mean they are observable, as this depends on the intrinsic amplitude the mode gets from the excitation mechanism.

\subsubsection{Mode suppression by internal magnetic fields}

\citet{Fuller2015} have shown that a strong magnetic field can suppress mixed modes in red giants. A similar phenomenon may be active in pulsating B stars \citep{Lecoanet2022}. It is thus worthwhile to ask what its effect could be for mode observability in HG BSGs and merger products.

As mentioned in Sect.\,\ref{sec:limitations}, we do not consider the presence of strong internal magnetic fields resulting from the merger process. We can estimate what the internal magnetic field strength would be if we assume a dipole magnetic field, that is, $B(r) = B_{\mathrm{surf}}(R_{\star}/r)^{3}$, with $B_{\mathrm{surf}}$ the surface magnetic field of our merger product \citep{schneider_long-term_2020}. $B_{\mathrm{surf}}$ can be estimated from the surface magnetic field of the MS merger product from \citet{schneider_stellar_2019} and assuming flux freezing, that is, $B_{\mathrm{MS}}R_{\mathrm{MS}}^{2} = B_{\mathrm{surf}}R_{\star}^{2}$, with $B_{\mathrm{MS}} = 9\times10^3\,$G and $R_{\mathrm{MS}} = 5\,\rsun$ the surface magnetic field and MS radius of the 3D MHD merger product from \citet{schneider_stellar_2019}. 

We can now compare this field strength with the critical magnetic field strength $B_{\mathrm{crit}} = \sqrt{\pi\rho/2}\omega^{2}r/N$, defined in \citet{Fuller2015} as the magnetic field strength above which the magnetic tension overcomes the buoyancy force. Here, $\omega$ is the angular mode frequency. We find that, using $\omega$ values from the range of mode frequencies predicted in this work, $B(r) < B_{\mathrm{crit}}$ in the outer g-mode cavity of the $6.0+2.4\,\msun$ merger product at the time it is compared with the $7.8\,\msun$ genuine single star, while in the inner g-mode cavity $B(r) > B_{\mathrm{crit}}$ (for modes with a period of 1 day, $B_{\mathrm{crit}} \sim 10^{4}\text{--}10^{7}\,\mathrm{G}$ in the inner g-mode cavity and $B_{\mathrm{crit}} \sim 5\times10^{4}\text{--}10^{7}\,\mathrm{G}$ in the outer g-mode cavity). Under these assumptions, we would expect the inner cavity g-modes to be suppressed by the magnetic field. However, this does not consider, among other uncertainties, that the magnetic field strength can be severely attenuated \citep{Quentin2018} or even expelled \citep{Braithwaite2017} when propagating through convective regions. Furthermore, \citet{Landstreet2007,Landstreet2008} and \citet{Fossati2016} show that the magnetic field strength in massive MS stars disappears faster than predicted from flux freezing alone.

\section{Discussion and conclusions}\label{sec:discussion}

Considering an ensemble of early Case B merger product models and genuine single HG stars, \citet{Bellinger2024} concluded that these two classes of objects cannot be distinguished based on their mean PSP values. However, these authors compared the $\Pi_{\ell = 1}$ values of all their models in the sets of merger products and genuine single stars (masses of $10\text{--}20\,\msun$) at all points during their BSG evolution simultaneously. Because of relatively large variations in $\Pi_{\ell}$ with evolutionary time and mass, which we also find in our models, there is a significant overlap between the ranges of $\Pi_{\ell}$ of both types of stars, which has led to the conclusion that they are indistinguishable based on $\Pi_{\ell}$. From our case-by-case comparison between models of these two classes at similar positions in the HRD, we conclude that the mean PSP value is consistently and significantly lower for merger products than for genuine single stars. We stress that for accurate predictions of the mean PSP value from the stellar structure, that is, $\Pi_{\ell}$, the integral in the denominator of Eq.\,(\ref{eq:buoyancy}) has to be evaluated over the proper g-mode cavity if more cavities are present in the model. In such cases, when the evanescent zone separating the cavities is substantial in size, the bulk of the g modes will be trapped in the inner cavity. Integrating over multiple cavities, as done in \citet{Bellinger2024}, leads to a deviation in the predictions of ${\lesssim}10\,\mathrm{s}$. This deviation is an order of magnitude smaller than the differences between the mean PSP values of post-MS merger products and genuine single HG stars found in our work but is comparable to the period precision of time series data from space missions.

We find that when a star has two g-mode cavities, which is the case for early Case B merger products at all masses considered in this work and genuine single HG stars with $M_{\star} \gtrsim 11.4\,\msun$, some inner-cavity g modes couple to outer-cavity g modes. This coupling leads to the formation of deep dips in the PSPs and can be used as a diagnostic to distinguish merger products from genuine single stars in the mass range roughly below $11.4\,\msun$. At higher masses, both the merger products and genuine single stars have two mode cavities, resulting in deep dips in their PSPs. In general, the appearance of deep PSP dips might not be unique to merger products since blue loop stars also have two mode cavities \citep{ostrowski_pulsations_2015}. Further detailed comparisons between genuine single HG stars, merger products, and blue loop stars should shed light on whether other differences can rule out blue loop stars in a population of BSGs.

From our initial results in Sect.\,\ref{sec:q_dep}, we conclude that the added mass fraction $f_{\mathrm{add}}$ has a relatively minor impact on the asteroseismic properties of our merger products. Depending on $f_{\mathrm{add}}$, we see that the merger products ignite He in their cores at different effective temperatures, leading to different stellar structures when the merger product occupies the same region of the HRD. With this exercise, we have explored how the pre-merger conditions might influence the asteroseismic properties of the merger product, even though $f_{\mathrm{add}}$ cannot be directly related to the mass ratio. Future exploration based on a grid of more complete merger models will allow us to determine whether the binary parameters at the time of merging are detectable in the asteroseismic properties of the merger product.

We find that the PSPs of early Case B merger products and genuine single HG stars are distinguishable when we ignore rotation, as well as when we include rotation at a level of $\Omega = 0.2\Omega_{\mathrm{c}}$. Taking rotation into account in the stellar oscillation computations for merger products is an important step forward for realistically predicting their asteroseismic properties. The modes most frequently observed are prograde sectoral dipole and quadrupole ($\ell = m = 1$ and $\ell = m = 2$, respectively) modes \citep{li_gravity-mode_2020,pedersen_internal_2021}. Furthermore, the fact that merger products are slow rotators is not firmly established, so asteroseismic predictions of fast-rotating merger products are warranted. Despite the many uncertainties on internal angular momentum transport, rotating equilibrium models with more realistic rotation profiles, such as those recently computed from 2-to-1D models by \citet{mombarg_two-dimensional_2024}, should be considered for such an exercise.

Finally, we conducted a set of preliminary tests to determine the observational potential of the modes predicted in our work. We find that it should, in principle, be possible to resolve period spacings down to $200\,$s, which is far below the mean PSP values predicted in this work, with five years of time series data. Depending on the depth of the deep PSP dips, which depends on how close the periods of modes in the inner and outer g-mode cavity lie, it might not be possible to resolve their minima at longer periods with less than five years of time series data. We stress again that a full assessment of the observability is plagued by uncertainties related to mode instability, the observable mode amplitudes at the stellar surface, and interior magnetic fields. Even though these uncertainties are pointed out here, 
we cannot meaningfully address mode observability as long as the mode excitation mechanisms remain incompliant with the observations as they are today and are unable to provide us with reliable predictions for the intrinsic mode amplitudes. 

\begin{acknowledgements}
We thank the anonymous referee for their valuable comments and suggestions, which have triggered further research and helped us to improve the presentation of our results. We thank J. Saling for the valuable insights from his Bachelor thesis titled ``Formation of Blue Supergiants in Stellar Mergers'' (University of Heidelberg, 2023). We thank T. Van Reeth, M. Michielsen, E. Bellinger, D. Bowman, E. Laplace, A. Noll, B. Bordad\'agua, Q. Copp\'ee, F. Ahlborn, C. Johnston, and R. Townsend (in no particular order) for the meaningful discussions and valuable comments and suggestions. Additional software used in this work includes \texttt{PyGYRE} \citep{townsend_pygyre_2020}, \texttt{PyMesaReader} \citep{wolf_py_mesa_reader_2017}, \texttt{MPI for Python} \citep{dalcin_mpi_2005,dalcin_mpi4py_2021}, \texttt{Astropy} \citep{astropy_collaboration_astropy_2013,astropy_collaboration_astropy_2018,astropy_collaboration_astropy_2022}, \texttt{NumPy} \citep{harris_array_2020} and \texttt{SciPy} \citep{virtanen_scipy_2020}. We used \texttt{Matplotlib} \citep{hunter_matplotlib_2007}, \texttt{matplotlib-label-lines} \citep{cadiou_matplotlib_2022}, and \texttt{mkipp} \citep{marchant_mkipp_2020} for plotting. The authors acknowledge support from the Klaus Tschira Foundation. This work has received funding from the European Research Council (ERC) under the European Union’s Horizon 2020 research and innovation programme (Starting Grant agreement N$^\circ$ 945806: TEL-STARS, Consolidator Grant agreement N$^\circ$ 101000296: DipolarSound, and Synergy Grant agreement N$^\circ$101071505: 4D-STAR). While funded by the European Union, views and opinions expressed are however those of the author(s) only and do not necessarily reflect those of the European Union or the European Research Council. Neither the European Union nor the granting authority can be held responsible for them. This work is supported by the Deutsche Forschungsgemeinschaft (DFG, German Research Foundation) under Germany’s Excellence Strategy EXC 2181/1-390900948 (the Heidelberg STRUCTURES Excellence Cluster).
\end{acknowledgements}

\FloatBarrier

\bibliographystyle{aa}
\bibliography{references}

\clearpage

\onecolumn

\begin{appendix}
\FloatBarrier
\section{Inclusion of slow rotation: TAR vs. perturbative inclusion of the Coriolis acceleration}\label{app:pert_vs_tar}
\begin{figure*}
    \sidecaption
    \includegraphics[width=12cm]{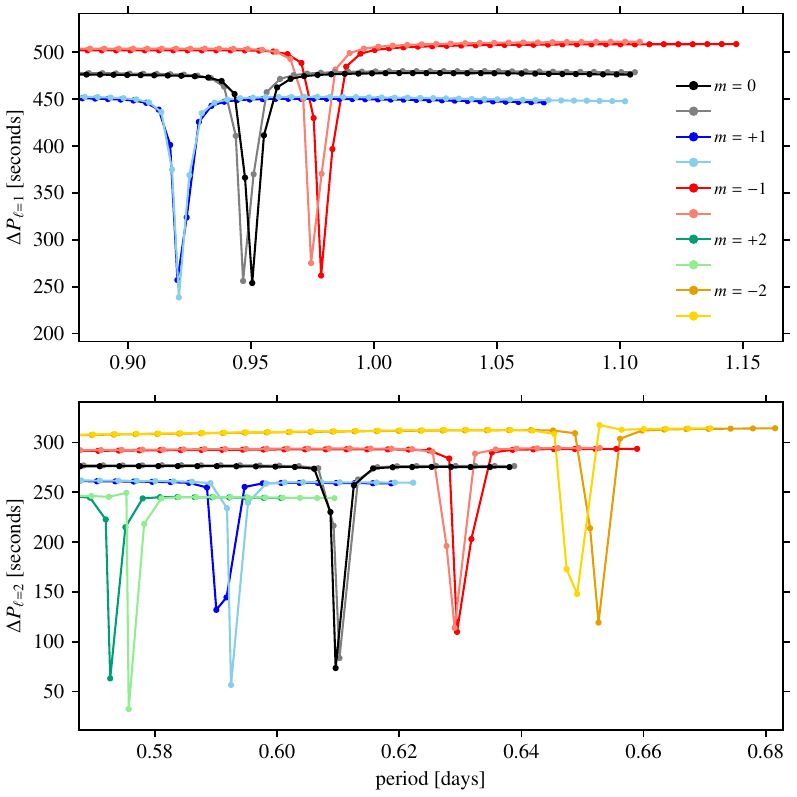}
    \caption{Period spacing patterns for a $6.0+2.4\,\mathrm{M}_{\odot}$ merger product with $\Omega = 0.2\Omega_{\mathrm{c}}$ in the inertial frame using the TAR and treating the Coriolis force as a perturbation. Light colours (grey, light blue, salmon, light green, and gold) indicate the results when the Coriolis force is treated as a perturbation. Dark colours (black, blue, red, bluish-green, and orange) show those for when the TAR is used. Panel (a) shows the period spacing patterns for $(\ell,\,m) = (1,\,-1)$, $(1,\,0)$, and $(1,\,+1)$ modes, and Panel (b) those for $(\ell,\,m) = (2,\,-2)$, $(2,\,-1)$, $(2,\,0)$, $(2,\,+1)$, and $(2,\,+2)$ modes.}
    \label{fig:tar_vs_pert}
\end{figure*}

We compute the spin parameters $s$ (see Sect.\,\ref{sec:diagnostics}) for the $(\ell,\,m) = (1,\,0)$ modes from the non-rotating calculations for the $6.0+2.4\,\msun$ merger product to determine whether we expect the modes to be super-inertial ($s < 1$) or sub-inertial ($s > 1$). We find values for $s$ between 0.02 ($n_{\mathrm{pg}} = -20$) and 0.13 ($n_{\mathrm{pg}} = -200$) when we assume $\Omega = 0.2\Omega_{\mathrm{c}}$, which means that the modes are super-inertial. However, the condition that $s \ll 1$, which is required for treating the Coriolis acceleration as a perturbation, is not strongly satisfied, especially for the higher-order modes. From the PSPs shown in Fig.\,\ref{fig:tar_vs_pert}, it is apparent that including the Coriolis acceleration as a perturbation leads to significant deviations from the solutions obtained using the TAR. These deviations are larger than the typical measurement errors for such modes, which are smaller than the plotted symbols in Fig.\,\ref{fig:tar_vs_pert} \citep{VanReeth2015a}. This shows that even though the oscillation modes considered in this work are super-inertial when $\Omega = 0.2\Omega_{\mathrm{c}}$, one should use the TAR instead of the first-order Ledoux perturbative approach (see \citealt{aerts_asteroseismic_2023} for details). This is especially true when the goal is to fit observed modes against theoretically predicted modes of such models (not in this work).

\FloatBarrier

\section{Propagation diagrams as a function of radial coordinate}\label{app:prop_radial}
\begin{figure*}
\centering
  \includegraphics[width=18cm]{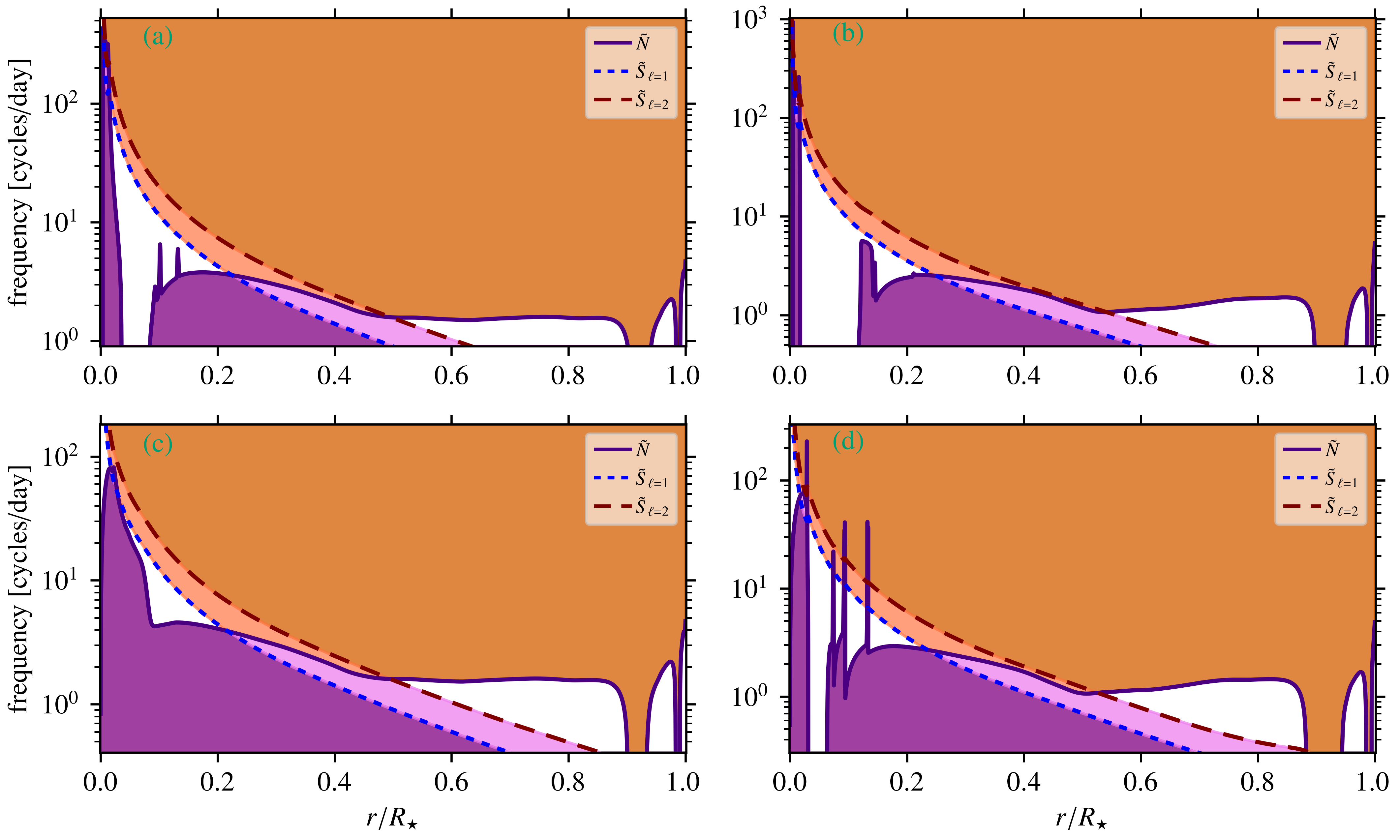}
     \caption{Propagation diagrams of the $6.0+2.4\,\msun$ merger product (a) and $7.8\,\msun$ (c) genuine single star, and the $9.0+6.3\,\msun$ merger product (b) and $13.6\,\msun$ genuine single star (d) from Fig.\,\ref{fig:propkip_compare1} and Fig.\,\ref{fig:propkip_compare2}, respectively, now as a function of the relative radial coordinate $r/R_{\star}$. The $\xi_{r}$ nodes are left out for clarity.}
     \label{fig:prop_radius}
\end{figure*}

\begin{figure*}
\centering
  \includegraphics[width=18cm]{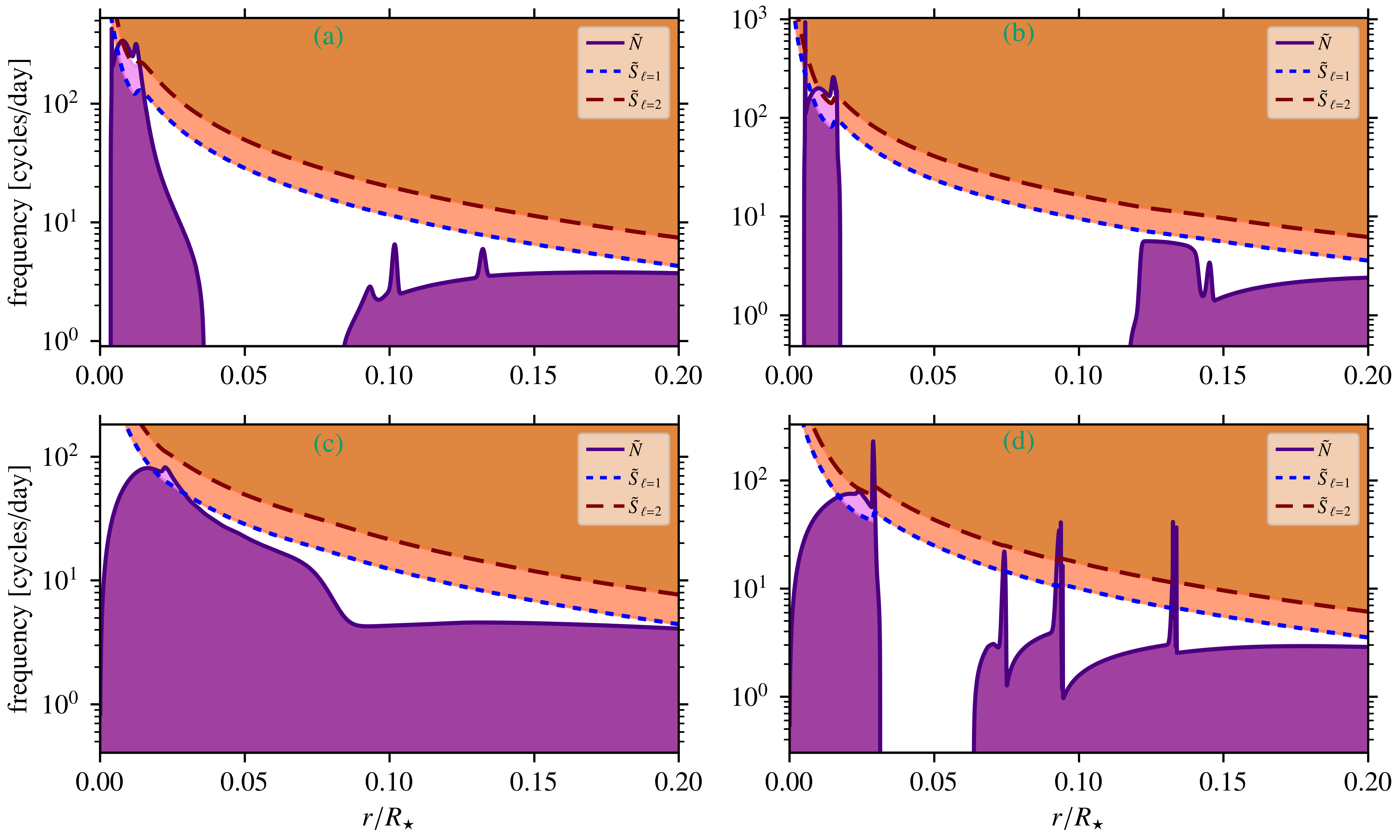}
     \caption{Same as Fig.\,\ref{fig:prop_radius}, now zoomed in on the region with $r/R_{\star} \leq 0.2$.}
     \label{fig:prop_radius_zoom}
\end{figure*}

In Fig.\,\ref{fig:prop_radius}, we show the propagation diagrams from Figs.\,\ref{fig:propkip_compare1} and \ref{fig:propkip_compare2} as a function of the relative radial coordinate $r/R_{\star}$ instead of relative mass coordinate $m/M_{\star}$. The $\xi_{r}$ nodes are left out for clarity. Figure \ref{fig:prop_radius_zoom} shows zoom-ins on the inner $20\%$ in relative radial coordinate of propagation diagrams in Fig.\,\ref{fig:prop_radius}.

\FloatBarrier

\section{Effect of semi-convection efficiency on ICZs}\label{app:effect_sc}

\begin{figure*}
    \sidecaption
    \includegraphics[width=9cm]{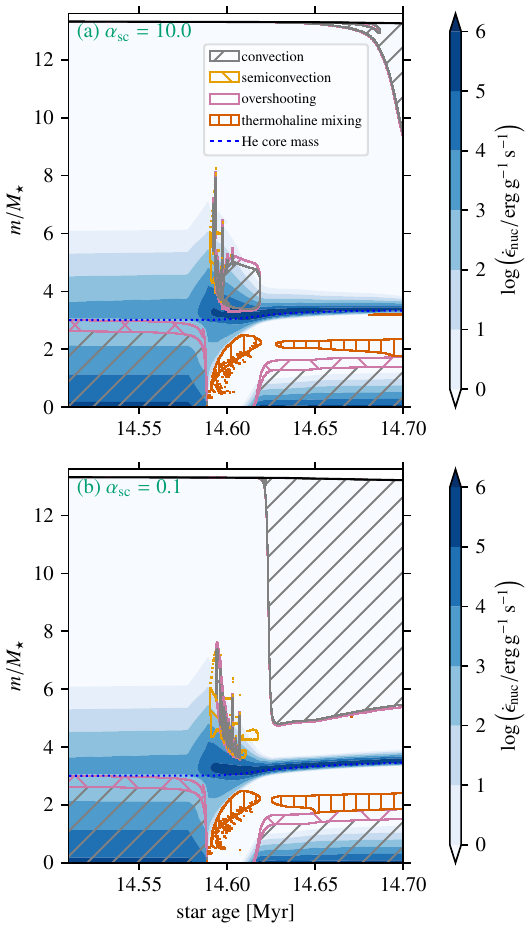}
    \caption{Kippenhahn diagram for a $13.6\,\msun$ genuine single star with intermediate convection zone for $\alpha_{\mathrm{sc}} = 10.0$ (a) and $\alpha_{\mathrm{sc}} = 0.1$ (b).}
    \label{fig:kipp_alphasc}
\end{figure*}

In Fig.\,\ref{fig:kipp_alphasc}a, we show a zoom-in on the Kippenhahn diagram of the $13.6\,\msun$ genuine single star during its HG evolution. In this zoom-in, the structure of the ICZ, which was first described in Sect.\,\ref{sec:compare2}, is more clearly shown. A more extended, non-uniform convection zone appears when the star arrives on the HG, which is responsible for the spiky features in the outer g-mode cavity of the $13.6\,\msun$ genuine single star (see Sect.\,\ref{sec:compare2}). Afterwards, a uniform ICZ appears, which persists until core-He ignition. As mentioned in Sect.\,\ref{sec:compare2} and demonstrated by \citet{kaiser_relative_2020} and \citet{sibony_impact_2023}, the extent and lifetime of these ICZs depend on the assumptions made for (semi-)convective mixing. To demonstrate this effect in our setup, we computed the same model with a semi-convection efficiency of $\alpha_{\mathrm{sc}} = 0.1$, shown in Fig.\,\ref{fig:kipp_alphasc}b. Also with lower values of $\alpha_{\mathrm{sc}}$, an ICZ appears. We note that the ICZ has a different morphology and lifetime. As shown in the main text of this work, ICZs are responsible for the appearance of deep dips in the PSPs of genuine single stars. The detection of these deep dips in the PSPs of genuine single stars with ICZ, if detectable at all, thus depends on the assumptions for (semi-)convection and on the time in the evolution that a genuine single HG star is observed. However, as demonstrated in Sect.\,\ref{sec:compare1} and \ref{sec:compare2}, even without an ICZ, merger products and genuine single stars are distinguishable based on their asymptotic period spacing values. 

\FloatBarrier

\section{Disentangling of dip structures in genuine single stars with an ICZ}\label{app:disentangle}

\begin{figure*}
    \sidecaption
    \includegraphics[width=9cm]{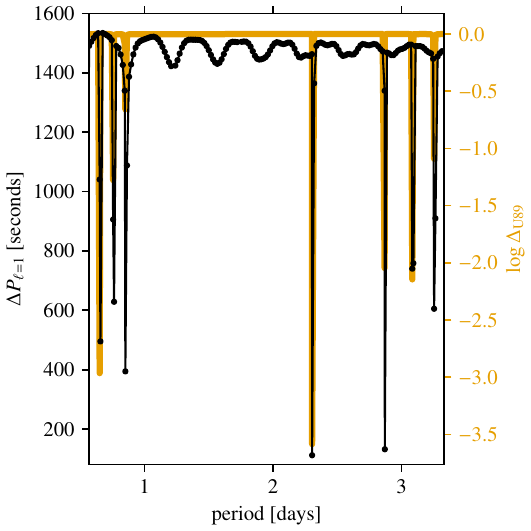}
    \caption{Same as Fig.\,\ref{fig:psp_DeltaUnno}, but for the $13.6\,\msun$ genuine single star and $(\ell,\,m) = (1,\,0)$ modes.}
    \label{fig:psp_DeltaUnno_single}
\end{figure*}
Fig.\,\ref{fig:psp_DeltaUnno_single} shows a part of the $13.6\,\msun$ genuine single star's PSP for $(\ell,\,m) = (1,\,0)$ modes without rotation, which is shown in full in Fig.\,\ref{fig:psps_compare2}. As in Fig.\,\ref{fig:psp_DeltaUnno}, we add the value of $\Delta_{\mathrm{U89}}$ (Eq.\,\ref{eq:delta_unno}) for each mode in the PSP. We see that only the deep, narrow dips have most of their kinetic energy in the outer g-mode cavity ($\Delta_{\mathrm{U89}} < 1$). Following the discourse from Sect.\,\ref{sec:deepdips}, these are the dips caused by mode coupling between inner- and outer-cavity g-modes. The more regular, quasi-periodic variation in the PSP involves only modes with most of their kinetic energy in the inner g-mode cavity ($\Delta_{\mathrm{U89}} \approx 1$). This confirms that the quasi-periodic variation of these modes in the PSP is caused by mode trapping in the inner g-mode cavity (see, e.g. \citealt{michielsen_probing_2021} for details on mode trapping).

\FloatBarrier

\section{Wave displacements and differential mode inertia for a selection of modes}\label{app:eigen}
\begin{figure*}
\centering
  \includegraphics[width=16cm]{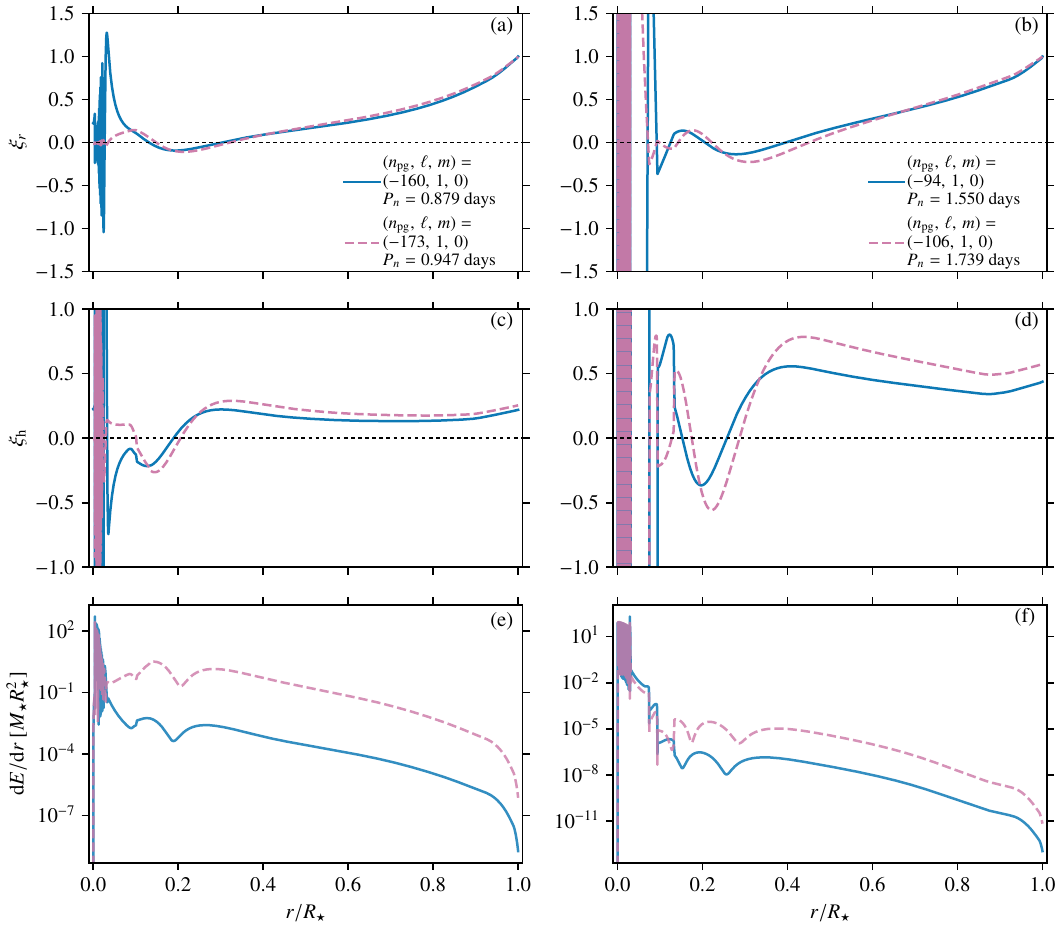}
     \caption{Wave displacement and differential mode inertia profiles for the $6.0+2.4\,\msun$ merger product (left column) and $7.8\,\msun$ genuine single star (right column) from nonadiabatic \texttt{GYRE} calculations for pure g modes in the long-period (low-frequency) regime. Panel (a)--(b) show the radial wave displacement $\xi_{r}(r)$, Panel (c)--(d) the horizontal wave displacement $\xi_{\mathrm{h}}(r)$, and Panel (e)--(f) the differential mode inertia $\mathrm{d}E/\mathrm{d}r$. The dashed pink lines in Panel (a), (c), and (e) show the aforementioned quantities for a mode in a deep PSP dip. The solid blue lines show those for a mode outside of a deep PSP dip, that is, for an inner-cavity g mode.}
     \label{fig:eigen_lower}
\end{figure*}

\begin{figure*}
\centering
  \includegraphics[width=16cm]{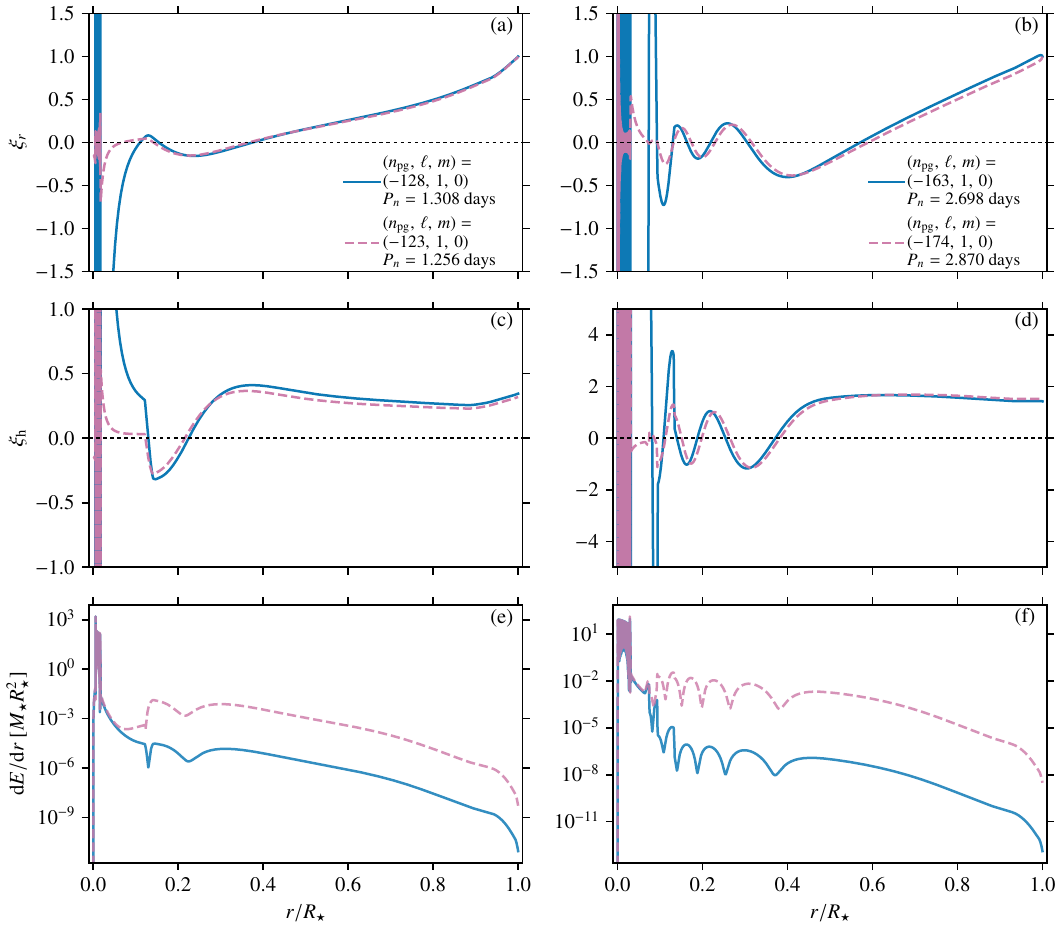}
     \caption{Same as Fig.\,\ref{fig:eigen_lower}, now for the $9.0+6.3\,\msun$ merger product (left column) and $13.6\,\msun$ genuine single star (right column). The dashed pink lines in all panels show the aforementioned quantities for a mode in a deep PSP dip. The solid blue lines show those for a mode outside of a deep PSP dip, that is, for an inner-cavity g mode.}
     \label{fig:eigen_higher}
\end{figure*}

\begin{figure*}
\centering
  \includegraphics[width=16cm]{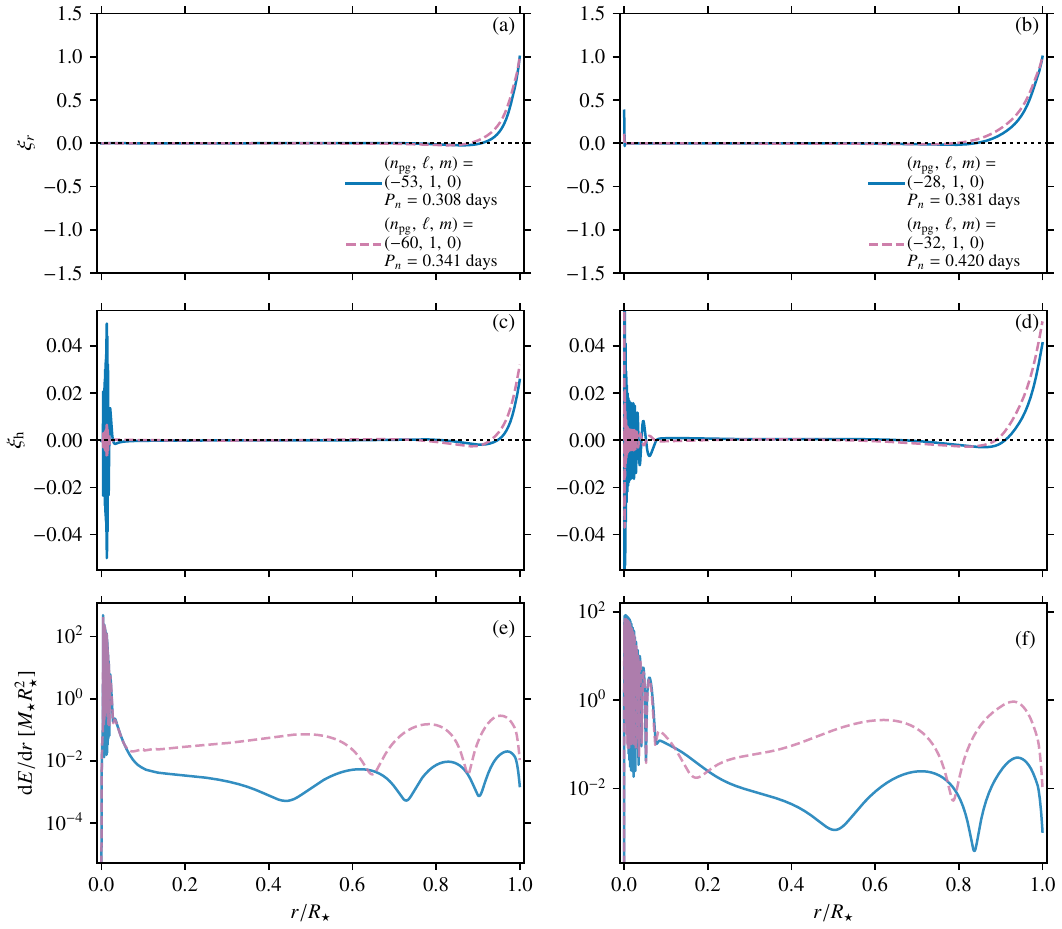}
     \caption{Same as Fig.\,\ref{fig:eigen_lower}, now for p-g mixed modes predicted for the $6.0+2.4\,\msun$ merger product (left column) and $7.8\,\msun$ genuine single star (right column). The dashed pink lines in all panels show the aforementioned quantities for a mode in a deep PSP dip. The solid blue lines show those for a mode outside of a deep PSP dip, that is, for an inner-cavity g mode. The radial orders $n_{\mathrm{pg}} = n_{\mathrm{p}} - n_{\mathrm{g}}$ of the p-g mixed modes in these plots are as follows: $n_{\mathrm{pg}} = -53 = 3 - 56$,  $n_{\mathrm{pg}} = -60 = 2 - 62$,  $n_{\mathrm{pg}} = -28 = 2 - 30$, and  $n_{\mathrm{pg}} = -32 = 1 - 33$.}
     \label{fig:eigen_mixed}
\end{figure*}

Figures \ref{fig:eigen_lower} and \ref{fig:eigen_higher} show the wave displacements $\xi_{r}(r)$, $\xi_{\mathrm{h}}(r)$ and the differential mode inertia $\mathrm{d}E/\mathrm{d}r$ (see Eq.\,3.139 in \citealt{aerts_asteroseismology_2010}) for a selection of long-period (high-frequency) pure g modes of the merger product and genuine single-star models described in Sect.\,\ref{sec:compare1} and \ref{sec:compare2}. We computed these mode properties with \texttt{GYRE}'s nonadiabatic setting to include nonadiabatic effects such as damping. Since we use the boundary conditions from \citet{unno_nonradial_1989}, the wave displacements at the surface behave as

\begin{equation}
    \frac{\xi_{\mathrm{h}}(r = R_{\star})}{\xi_{r}(r = R_{\star})} \simeq \frac{GM_{\star}}{R_{\star}^{3}\omega^2}\,.
\end{equation}
Using this relation, we normalised the wave displacements such that $\xi_{r}(r = R_{\star}) \equiv 1$. In Fig.\,\ref{fig:eigen_mixed}, we show the equivalent plots for p-g mixed modes (shorter period, higher frequencies) for the $6.0+2.4\,\msun$ merger product and $7.8\,\msun$ genuine single star.

\FloatBarrier
\end{appendix}

\end{document}